\pdfoutput=1

\documentclass[twocolumn,showpacs,amssymb,aps,nofootinbib,floatfix,superscriptaddress]{revtex4-1}

\usepackage{epsfig}
\usepackage{float}
\usepackage{amssymb}
\usepackage{hyperref}

\begin{document}

\title{Longitudinal decorrelation measures of flow magnitude and event-plane angles \\in ultra-relativistic nuclear collisions}

\author{Piotr Bo\.zek}
\email{Piotr.Bozek@fis.agh.edu.pl}
\affiliation{AGH University of Science and Technology, Faculty of Physics and
Applied Computer Science, al. Mickiewicza 30, 30-059 Cracow, Poland}

\author{Wojciech Broniowski}
\email{Wojciech.Broniowski@ifj.edu.pl}
\affiliation{The H. Niewodnicza\'nski Institute of Nuclear Physics, Polish Academy of Sciences, 31-342 Cracow, Poland}
\affiliation{Institute of Physics, Jan Kochanowski University, 25-406 Kielce, Poland}

\date{9 November 2017}

\begin{abstract}
We discuss the forward-backward correlations of harmonic flow in Pb+Pb collisions at the CERN Large Hadron Collider (LHC), 
applying standard multibin measures, as well as proposed here new measures. 
We illustrate the methods with 
hydrodynamic model simulations based on event-by-event initial conditions from the wounded quark model with 
asymmetric rapidity emission profiles.
Within the  model we examine 
independently the event-plane angle and the flow magnitude decorrelations. 
 We find specific hierarchy between various flow decorrelation measures and confirm certain factorization relations.
We find qualitative agreement of the model and the data from the ATLAS and CMS Collaborations.
\end{abstract}



\keywords{ultra-relativistic nuclear collisions, event-by-event fluctuations,forward-backward  harmonic flow correlations}

\maketitle

\section{Introduction}

In recent years, longitudinal correlations of harmonic flow have been the object of an active study in the field of ultra-relativistic nuclear 
collisions. The importance of these investigations stems from the fact that the obtained correlation measures, related to the harmonic flow which 
is the key physical phenomenon of the underlying collective dynamics, probe the fluctuations in the early times of the evolution of the system, and as such 
pose severe and challenging constraints to models of the initial stages of the collisions.

It is widely believed that the harmonic flow seen in momenta distributions of the produced hadrons in the transverse 
plane originates from the eccentricity generated in the initial state via the mechanism of the shape-flow transmutation.
Since the initial state is approximately boost-invariant, at least in the pseudorapidity acceptance of the current experiments, 
one expects the flow pattern to be very similar not too far away from mid-rapidity. This is indeed the case, as can be  
inferred from the recent data from the CMS~ \cite{Khachatryan:2015oea} and ATLAS~\cite{Huo:2017hjv,Aaboud:2017tql} Collaborations 
at the LHC, as well as from a very recent analysis at the BNL Relativistic Heavy-Ion Collide (RHIC)~\cite{NieTalk}.

However, the expected large correlation of flow properties in distant pseudorapidity bins 
is diminished by the presence of fluctuations in the early dynamics. These fluctuations may be of different 
origin, and affect both the transverse and the longitudinal event-by-event distribution of the entropy density. 
Studies of appropriate measures in different collision systems at different centralities and at different collision energies may help to 
understand the pertinent mechanisms of early fluctuations.

The {\em torque} effect, proposed  in~\cite{Bozek:2010vz}, has fluctuating angles of the event-plane orientation as a function of pseudorapidity, 
caused by fluctuations of the longitudinal distribution of sources. As described in Sect.~\ref{sec:model}, these fluctuations are induced by 
sources\footnote{By operational definition, a source is an early object which deposits entropy in the fireball. Typical model 
realizations are wounded nucleons, wounded quarks, or gluonic hot-spots.} of a random character,
which have asymmetric emission profiles in rapidity. They deposit entropy preferentially in the forward or backward direction, 
depending if they come from one or the other colliding nucleus.
As a result, the forward and backward event-plane angles are decorrelated; 
the effect is not large, 10-20 degrees over a few units of pseudorapidity, 
but clearly visible with the accurate experimental data.
The idea was further developed in~\cite{Huo:2013qma,Jia:2014vja,Jia:2014ysa}, and the basic concept was confirmed with the 
experimental findings~\cite{Khachatryan:2015oea,Huo:2017hjv,Aaboud:2017tql}.
First hydrodynamic studies with the wounded-nucleon model~\cite{Bialas:1976ed} initial 
conditions~\cite{Bozek:2015tca} or with the AMPT~\cite{Lin:2004en} initial conditions~\cite{Pang:2015zrq} led to qualitative, or semi-quantitative, agreement
for the Pb+Pb collisions at the LHC energies.

An analogous phenomenon is
the transverse momentum ($p_T$) {\em factorization breaking} proposed in~\cite{Gardim:2012im,Heinz:2013bua,Kozlov:2014fqa}, 
where the event-plane orientation decorrelates as a function of $p_T$ (in the same pseudorapidity bin). 
This phenomenon was experimentally confirmed for Pb+Pb and p+Pb by ALICE~\cite{Acharya:2017ino},
and for p+Pb by the ATLAS~\cite{Aad:2014lta} Collaborations.

With the results for the torque effect for p+Pb collisions from~\cite{Acharya:2017ino} it became clear that the fluctuation 
mechanism is more subtle. Extra fluctuations from string breaking~\cite{Bozek:2015bna,Broniowski:2015oif,Monnai:2015sca,Sakai:2017rfi},
glasma~\cite{Schenke:2016ksl}, or decelerating strings~\cite{Shen:2017bsr}
were investigated. Yet, a uniform physical model of flow decorrelation, working for different 
collision systems (Pb+Pb, p+Pb), centralities, and rank of the flow harmonic, is to our knowledge not yet constructed.

\begin{figure}
\begin{center}
\includegraphics[angle=0,width=0.4 \textwidth]{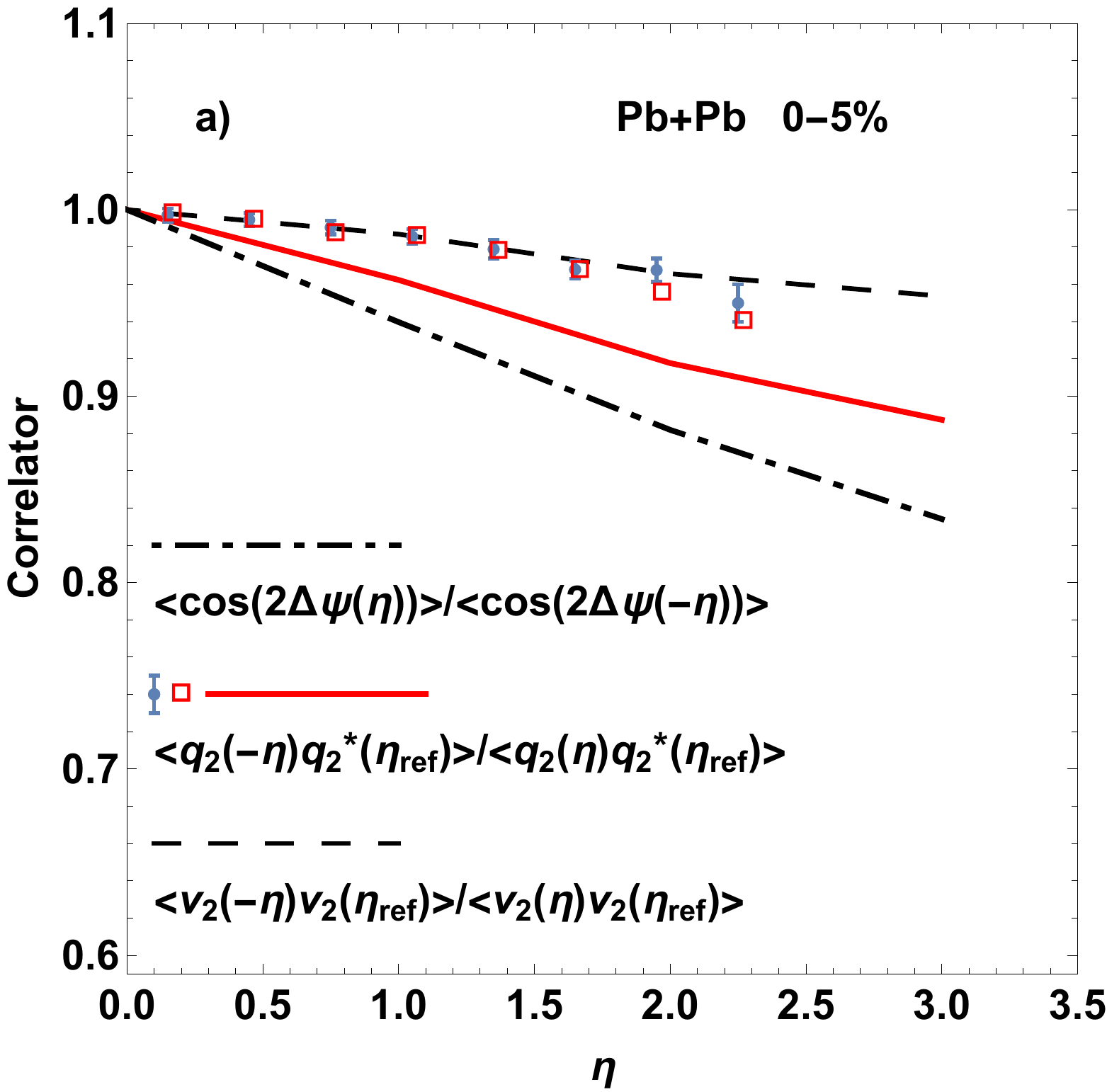}
\vspace{4mm}  

\includegraphics[angle=0,width=0.4 \textwidth]{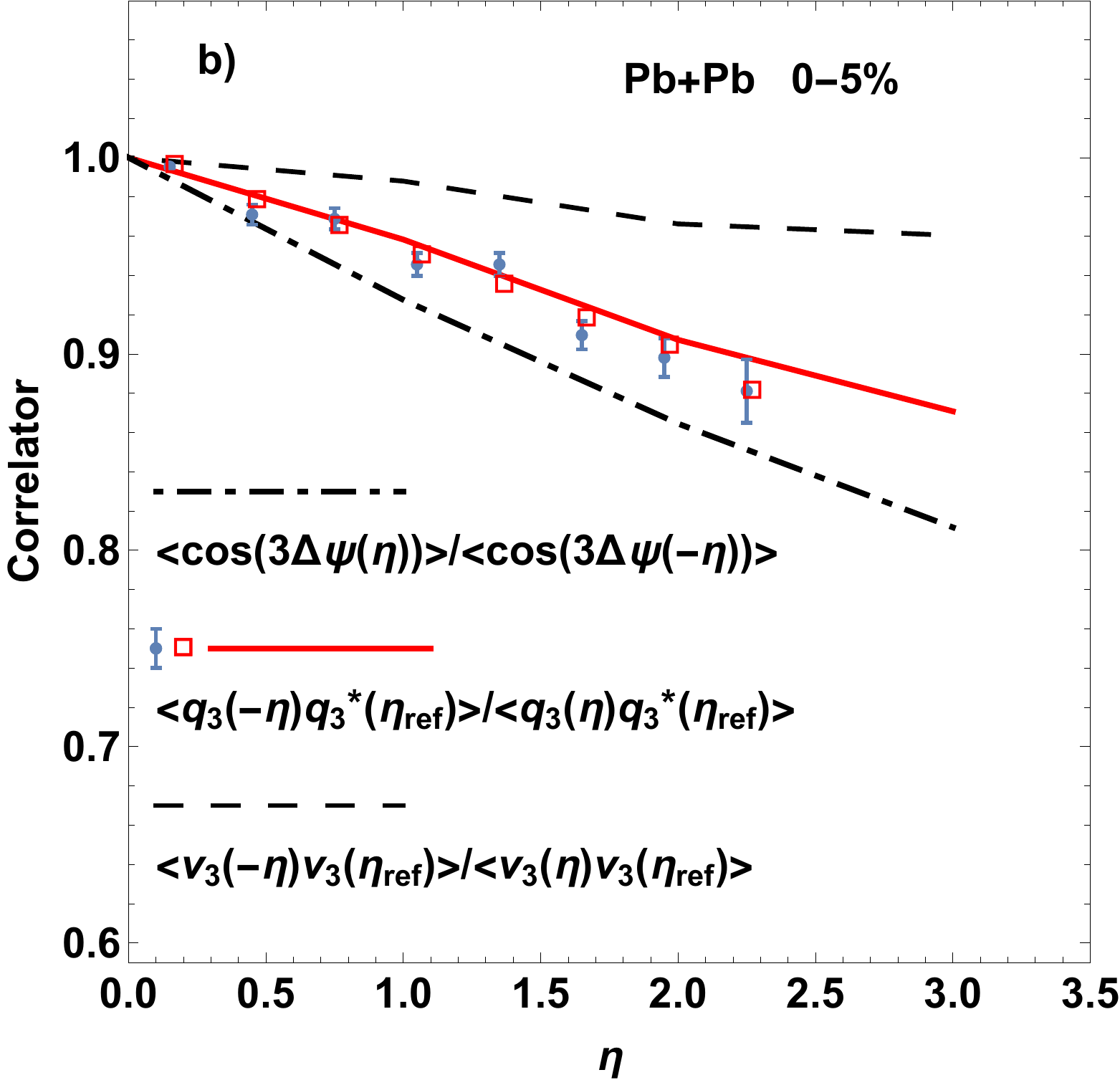}  
\end{center}
\vspace{-5mm}
\caption{3-bin correlator for the flow angle from Eq.~(\ref{eq:dpsi}) (dot-dashed line), for the flow magnitude from Eq.~(\ref{eq:dv2mag}) 
(dashed line), and for the harmonic flow measure from Eq.~(\ref{eq:dvn})  (solid line for the model calculation described in the text. Diamonds 
indicate the 
CMS Collaboration data \cite{Khachatryan:2015oea}, and the open squares the ATLAS Collaboration
data \cite{Aaboud:2017tql}). Panels (a) and (b) show the results for the second- and third-order harmonic
flow, respectively. Pb-Pb collisions  for centrality 0-5\%.
\label{fig:vv05}} 
\end{figure}

\begin{figure}
\begin{center}
\includegraphics[angle=0,width=0.4 \textwidth]{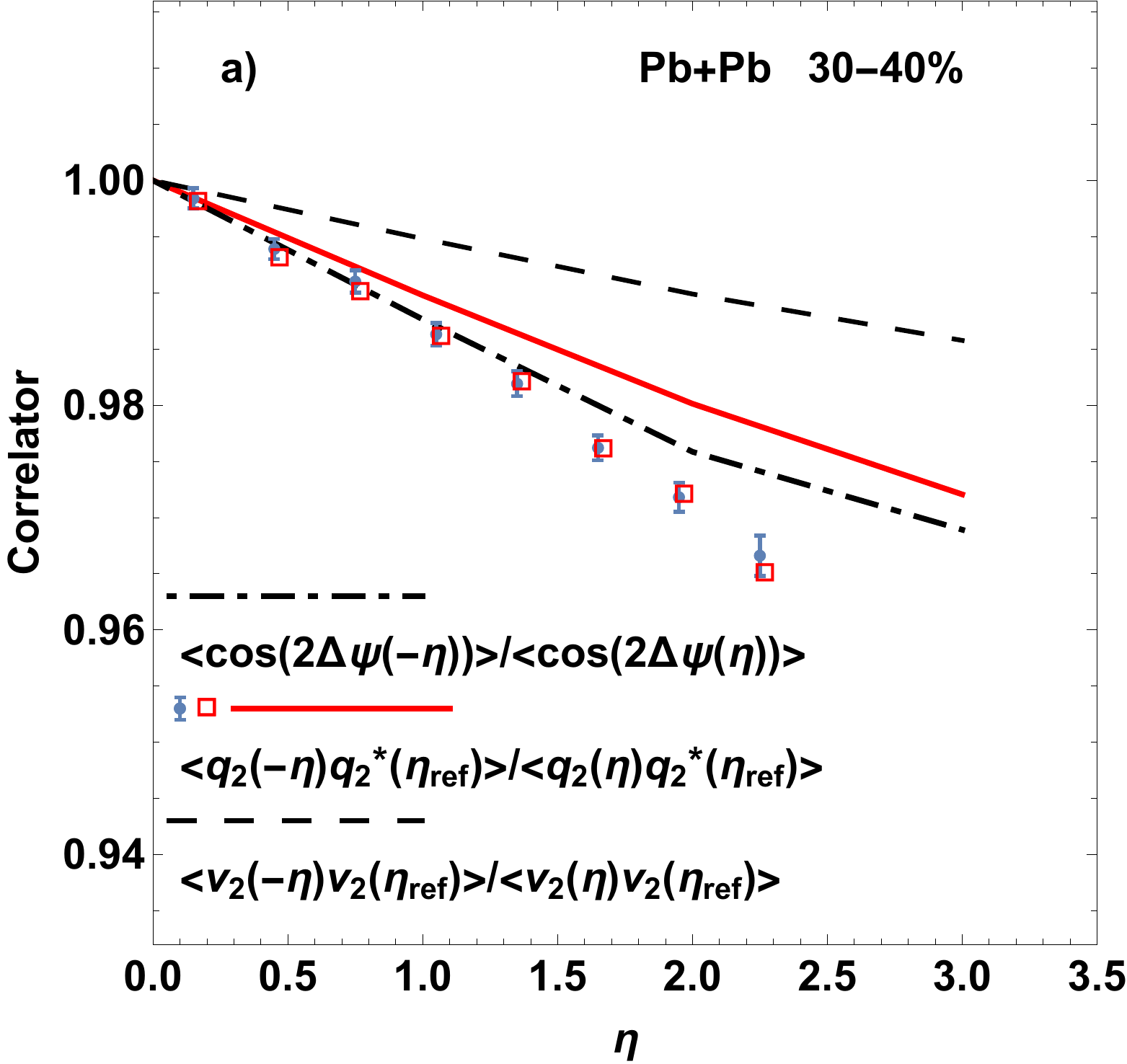}  
\vspace{4mm}

\includegraphics[angle=0,width=0.4 \textwidth]{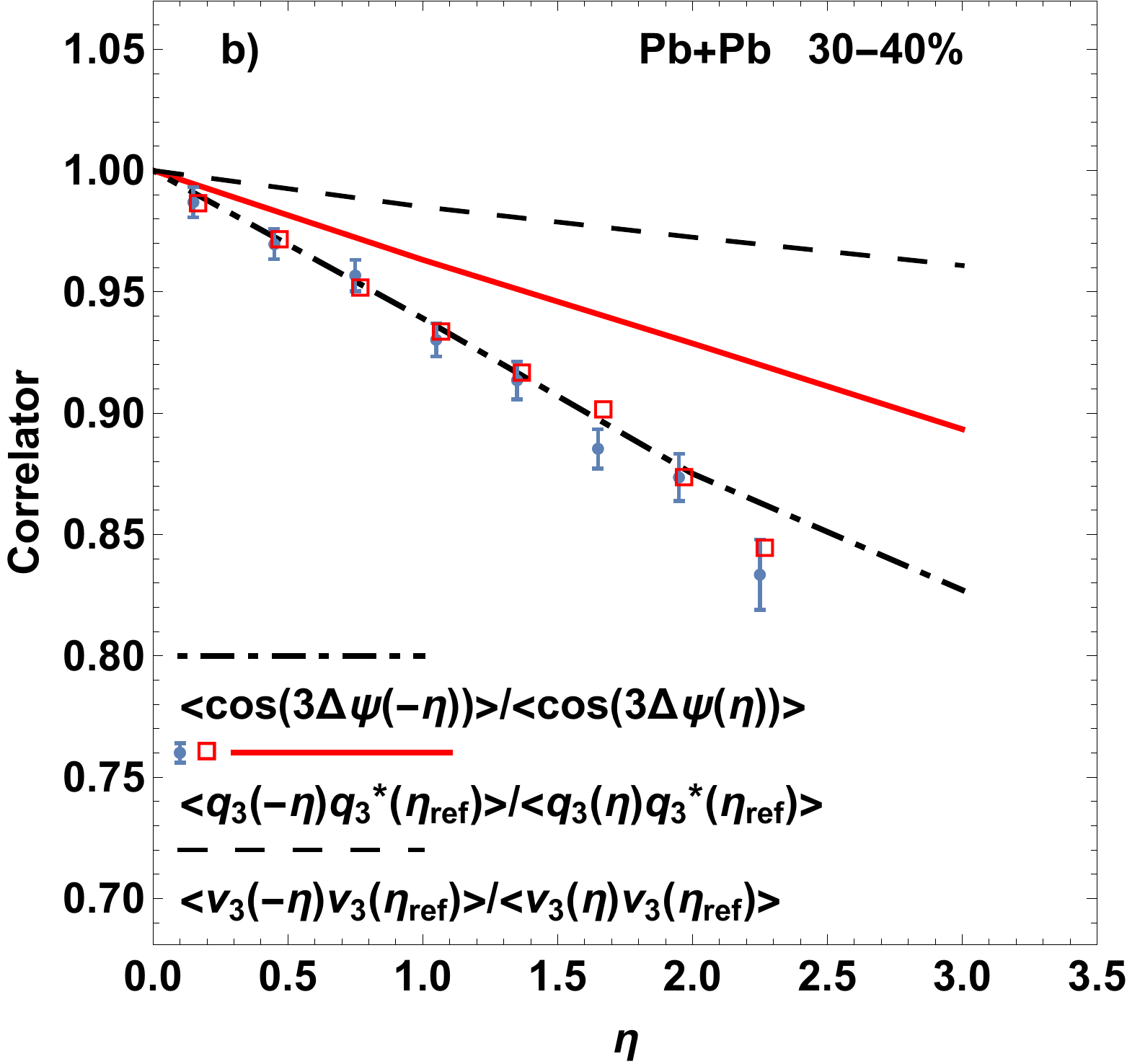}  
\end{center}
\vspace{-5mm}
\caption{Same as Fig. \ref{fig:vv05} but for centrality 30-40\%.
\label{fig:vv3040}} 
\end{figure}

The purpose of this paper is to discuss the methodology of the longitudinal event-by-event fluctuations of the 
harmonic flow in Pb+Pb collisions at the Large Hadron Collider (LHC),
with a focus on the interplay between the angle and flow magnitude correlations.
In addition to by-now standard multibin decorrelation measures, we investigate new measures, sensitive 
to the correlations of  the event-plane torque angle  and the flow magnitude.
We illustrate the techniques with a 3+1D hydrodynamic model run event-by-event. The outcome is specific hierarchy between various decorrelation measures. 
We also confirm factorization relations between the flow angle and magnitude decorrelations in specific correlators, first proposed in~\cite{Aaboud:2017tql}. 
Our generic predictions and new decorrelation measures can be useful for future data analyses. An interesting possibility in this avenue has been explored in~\cite{Ke:2016jrd}, where a procedure of reverse engineering for the 
initial conditions from the final correlations is performed.   Data and 
simulation on different correlators of flow harmonics  would provide
 important additional data for the Bayesian fitting procedure.

\section{Model \label{sec:model}}

We use a 3+1 dimensional viscous hydrodynamic model to describe the 
evolution of the dense system formed in the early stage of  
the collision~\cite{Bozek:2009dw,Schenke:2010rr}.
The initial conditions are provided by the Glauber Monte Carlo model with quark degrees of freedom~\cite{Rybczynski:2013yba}, which 
is phenomenologically successful in describing the multiplicity distributions over a wide range of reactions~\cite{Bozek:2016kpf}.
The expansion of the fluid stops at the freeze-out temperature $150$~MeV, with subsequent statistical emission of particles from the 
freeze-out hypersurface modeled with {\tt THERMINATOR}~\cite{Chojnacki:2011hb}.

\begin{figure}
\begin{center}
\includegraphics[angle=0,width=0.4 \textwidth]{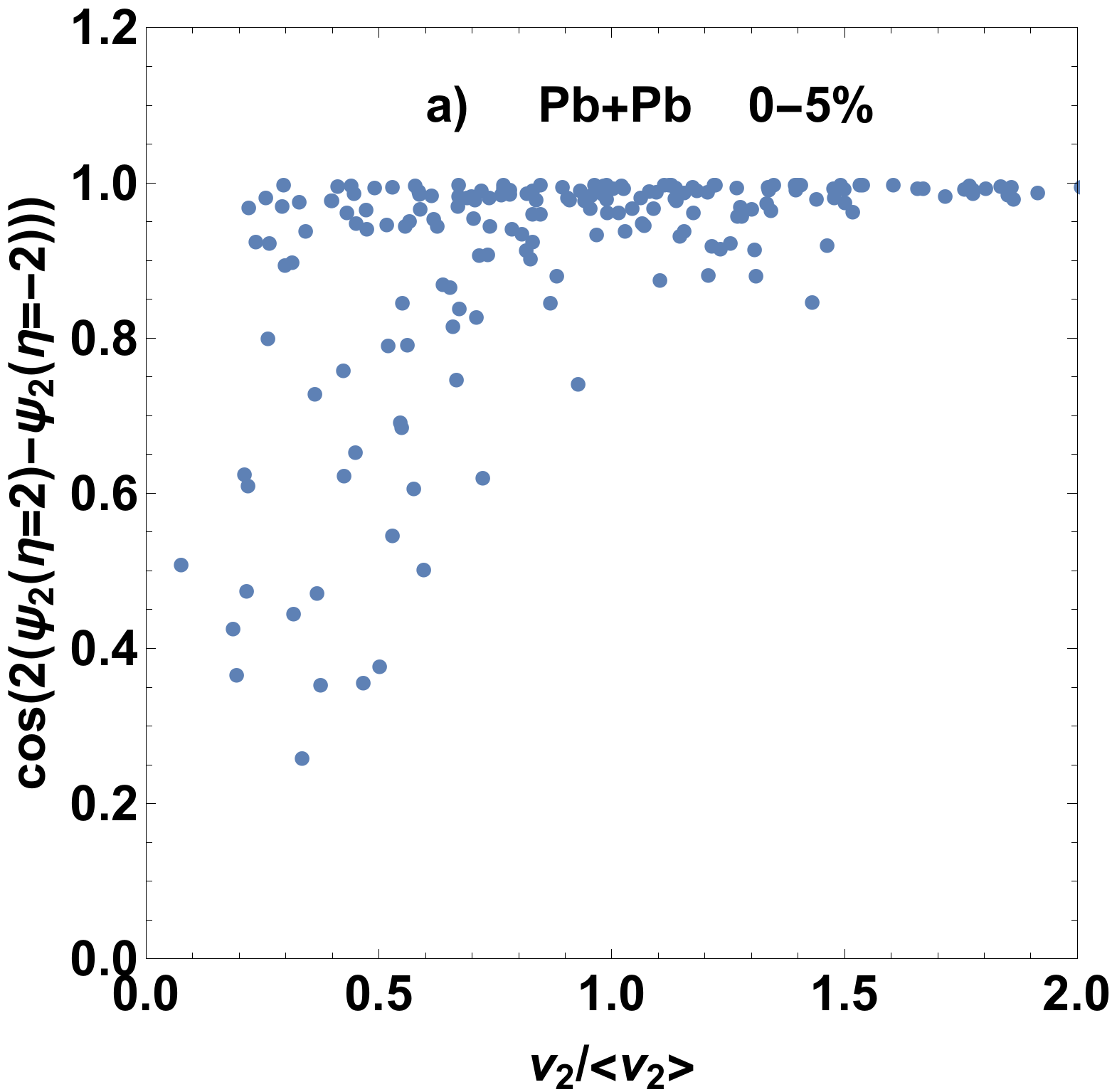}  
\vspace{4mm}

\includegraphics[angle=0,width=0.4 \textwidth]{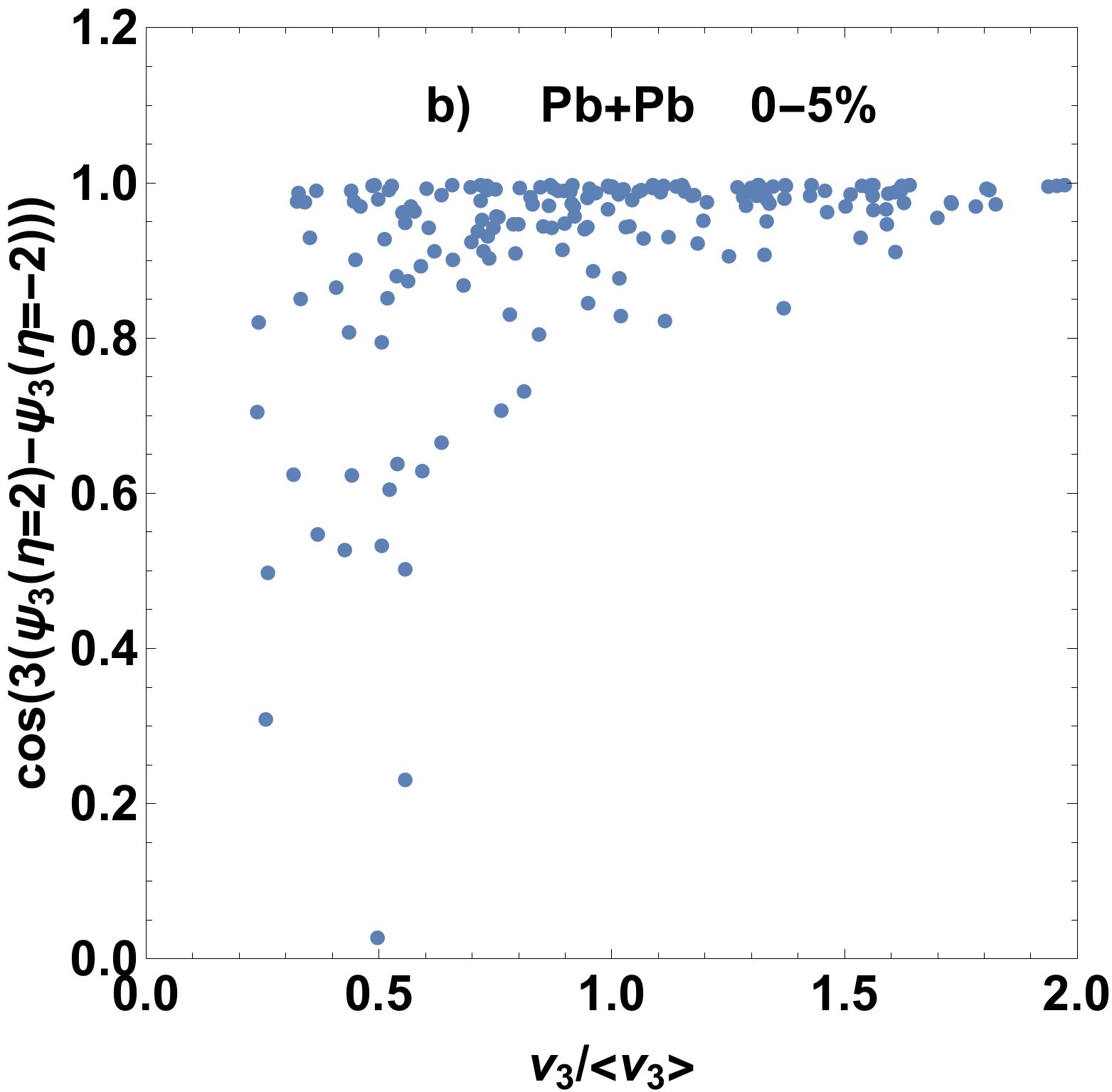}  
\end{center}
\vspace{-5mm}
\caption{Scattered plot of  $\cos\left(n\left(\Psi_n(\eta=2)-\Psi_n(\eta=-2)\right)\right)$ versus 
the flow magnitude $v_n$ for the hydrodynamic events, for the second- (a) and third-order (b) harmonic flow in Pb-Pb collisions 
at centrality 0-5\%. \label{fig:scat05}} 
\end{figure} 

\begin{figure}
\begin{center}
\includegraphics[angle=0,width=0.4 \textwidth]{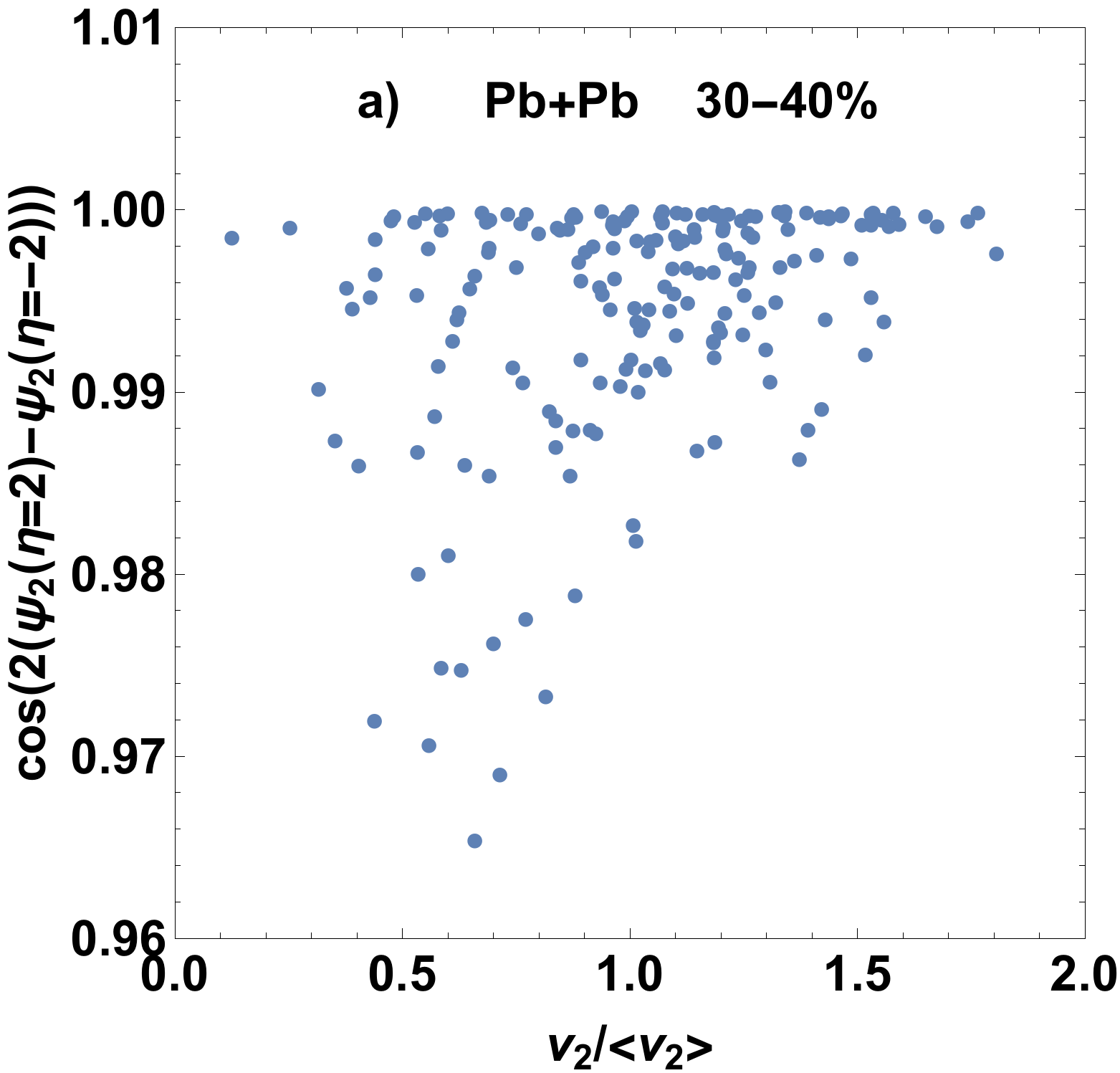}  
\vspace{4mm}

\includegraphics[angle=0,width=0.4 \textwidth]{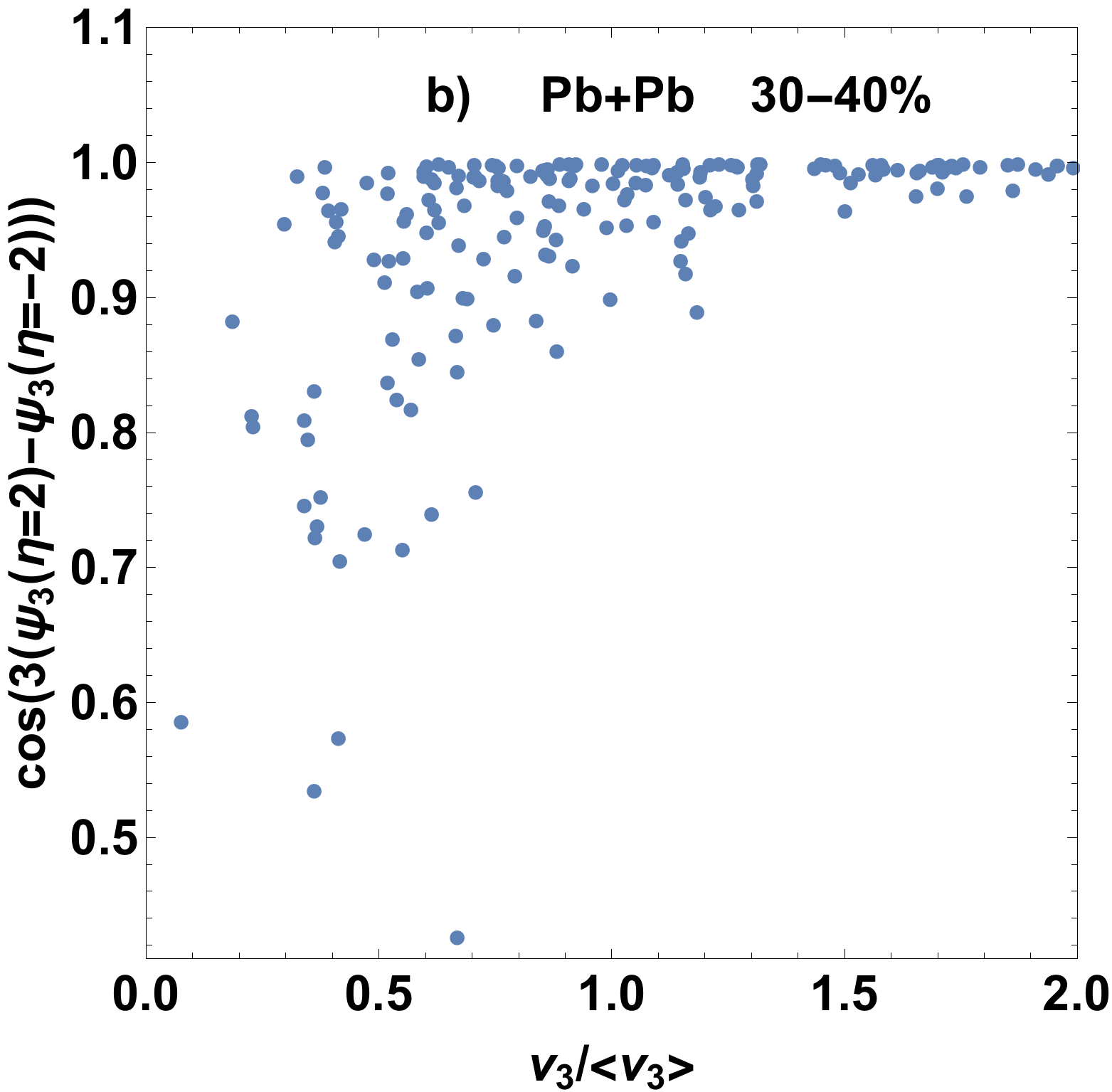}  
\end{center}
\vspace{-5mm}
\caption{Same as Fig. \ref{fig:scat05} but for centrality 30-40\%.
\label{fig:scat3040}} 
\end{figure} 

The fluctuations of the forward and backward going participant quarks lead to
fluctuations in space-time rapidity $\eta_\parallel = \frac{1}{2} \log \frac{t+z}{t-z}$ of the initial entropy density, which can be written as
\begin{equation}
s(x,y,\eta_\parallel)=\sum_{i=1}^{N_+} g_i(x,y) f_+(\eta_\parallel) + \sum_{i=1}^{N_-} g_i(x,y) f_-(\eta_\parallel)   \ . \label{eq:entropy}
\end{equation} 
The sums run over  $N_+$ right- and $N_-$ left-going participants at a transverse position $(x_i,y_i)$,
and each source contributes a Gaussian-smeared term in the transverse plane,
\begin{equation}
g_i(x,y)= \kappa  \exp\left( - \frac{(x-x_i)^2+(y-y_i)^2}{2\sigma^2}\right) .
\end{equation} 
The longitudinal profiles are asymmetric,
\begin{equation}
f_{\pm}(\eta_\parallel)= \frac{y_{\rm beam}\pm \eta_\parallel}{2 y_{\rm beam}} H(\eta_\parallel)\  \mbox {for } \ |\eta_\parallel|<y_{\rm beam}
\label{eq:lprof}
\end{equation}
hence the  entropy is deposited preferably  in the direction of motion of a given participant~\cite{Bialas:2004su}.

This simple model (with a factorized transverse-longitudinal distribution of Eq.~(\ref{eq:entropy}) where only the transverse component fluctuates) 
leads to torque-like fluctuations of the orientation of the principal axes of 
eccentricities at different space-time rapidities~\cite{Bozek:2010vz}.

Our simulations, carried out  within the above model, are as follows:
For each hydrodynamic event we generate $200$-$800$ {\tt THERMINATOR} events,
depending on centrality. These events are combined together such that for each
hydrodynamic event the flow pattern can be reconstructed with good accuracy.
This simple {\em cumulative-event} procedure~\cite{Bozek:2015bha} removes any non-flow correlations
(in the case of {\tt THERMINATOR}, those coming from resonance decays).
The reconstruction of the flow vectors allows us to check how the experimental
observables are related to the actual flow angle and magnitude decorrelation in the model.

\section{Decorrelation of the flow event-plane angles and magnitudes \label{sec:decor}}

The basic object in constructing the flow measures is the flow vector, defined in each event as
\begin{equation}
q_n(\eta)=\frac{1}{m}\sum_{k=1}^{m} e^{i n \phi_k} \equiv v_n(\eta) e^{i n \Psi_n(\eta)} \ ,
\end{equation}
where the sum runs over the $m$ hadrons  in a pseudorapidity interval around $\eta$, $\phi_k$ is the azimuth of a hadron's 
motion, $v_n$ is the magnitude of harmonic flow of order $n$ in the interval, and $\Psi_n$ is the corresponding event-plane angle.

The decorrelation of the harmonic flow in pseudorapidity can be quantified
using the two-particle correlation
\begin{equation}
V_{n\Delta}(\eta_1,\eta_2)= \langle q_n(\eta_1) q_n^\star(\eta_2)
\rangle \ , \label{eq:cv}
\end{equation}
where  $\langle \dots \rangle$ denotes the average over events.
The function (\ref{eq:cv}) is a measure of the correlation of
the flow vector of harmonic order $n$ at pseudorapidities $\eta_1$ and $\eta_2$. For small rapidity separation of the two bins
($\Delta \eta <1$)), the correlation 
function involves a significant contribution from the non-flow correlations~\cite{Bozek:2010vz,Bozek:2015bha}.

The ratio of the correlation functions at two different (but sufficiently large) pseudorapidity separations is a clever measure of flow decorrelation, 
which largely reduced the non-flow correlations~\cite{Khachatryan:2015oea}. It involves three pseudorapidity bins and is defined as
\begin{equation} 
r_n(\eta) \equiv r_{n|n;1}(\eta)=\frac{\langle q_n(-\eta)q_n^\star(\eta_{\rm ref})\rangle}{\langle  q_n(\eta)q_n^\star(\eta_{\rm ref}) \rangle}.
\label{eq:dvn}
\end{equation}
The reference bin at $\eta_{\rm ref}$ is taken sufficiently far in the forward or backward pseudorapidity region, such that both $|\eta_{\rm ref}-\eta|$ 
and $|\eta_{\rm ref}+\eta|$ are large enough to suppress the non-flow correlations. In our simulations, following the experimental 
set up~\cite{Khachatryan:2015oea}, we take the reference bin as
$4.4<\eta_{\rm ref}<5$. We also symmetrize between the forward and backward pseudorapidities to increase statistics.

One should recall that an important feature of measure (\ref{eq:dvn}), shared with other measures used to quantify 
event-by-event fluctuations, is the cancellation
of the random fluctuations 
resulting from a finite number of hadrons in the pseudorapidity bins. That way the unfolding 
of the trivial statistical component from hadronization is effectively carried out.

The deviation of the  correlator $r_{n|n;1}$ from unity is a measure
of a combined decorrelation of the flow angle $\Psi_n$  and of  the flow magnitude $v_n$ at different rapidities~\cite{Aaboud:2017tql}, 
since
\begin{equation}
r_{n|n;1}(\eta)= \frac{\langle v_n(-\eta)v_n(\eta_{\rm ref}) \cos\left[ n (\Psi_n(-\eta)-\Psi_n(\eta_{\rm ref}))\right] \rangle}
{\langle v_n(-\eta)v_n(\eta_{\rm ref}) \cos\left[ n (\Psi_n(-\eta)-\Psi_n(\eta_{\rm ref}))\right] \rangle} \ .
\end{equation} 

It has been noticed in the numerous model 
calculations~\cite{Bozek:2015tca,Lin:2004en,Pang:2015zrq,Bozek:2015bna,Broniowski:2015oif,Sakai:2017rfi,Schenke:2016ksl,Monnai:2015sca,Shen:2017bsr,Ke:2016jrd} 
and in the experimental data \cite{Khachatryan:2015oea,Huo:2017hjv,Aaboud:2017tql} that
the flow decorrelation is larger for centralities where the flow
magnitude is small (such as the central collisions for $v_2$).
%
To gain more insight, in a model one can study separately the decorrelation of the magnitude and of the flow angle, and we will carry out 
this analysis in our hydrodynamic framework applying the cumulative-event procedure to increase statistics and remove the non-flow effects.
The flow angle decorrelation can be calculated from the event-by-event orientation of $\Psi_n$ as a function of rapidity,
\begin{equation}
r_n^{\Psi}(\eta)=\frac{\langle \cos\left[ n\left(\Psi_n(-\eta)-\Psi_n(\eta_{\rm ref})\right)\right] \rangle}
{\langle \cos\left[n\left(\Psi_n(\eta)-\Psi_n(\eta_{\rm ref})\right)\right]\rangle}.
\label{eq:dpsi}
\end{equation}
Analogously, the flow magnitude decorrelation can be measured as
\begin{equation}
r_n^{v}(\eta)=\frac{\langle v_n(-\eta)v_n(\eta_{\rm ref})\rangle}{\langle  v_n(\eta)v_n(\eta_{\rm ref}) \rangle} \ .
\label{eq:dv2mag}
\end{equation}

In Figs. \ref{fig:vv05} and \ref{fig:vv3040} we show the results for the 
decorrelations of the flow angle, flow magnitude, and flow vector for the second- and third-order
harmonic flow in Pb+Pb collisions at the LHC. In can be noticed that the correlator for the torque angle shows the
largest deviation from $1$, i.e., the decorrelations of the angle is significantly 
larger than the magnitude decorrelation. 
It is even more striking to find that  the decorrelation of the flow angle $r_n^\Psi$
is larger than the flow decorrelation $r_2$, although the latter one 
measures the combined decorrelation of both the flow angle and the magnitude.
As we will discuss in detail shortly, the effect comes from the fact that the measure $r_n$ averages $\cos(n \Delta \Psi )$ 
weighted with $v_n^2$, and these variables are strongly correlated. This effect is especially important for $v_3$ and for $v_2$ in 
central collisions, where the relative fluctuations of $v_n$ are large.

\begin{figure}
\begin{center}
\includegraphics[angle=0,width=0.4 \textwidth]{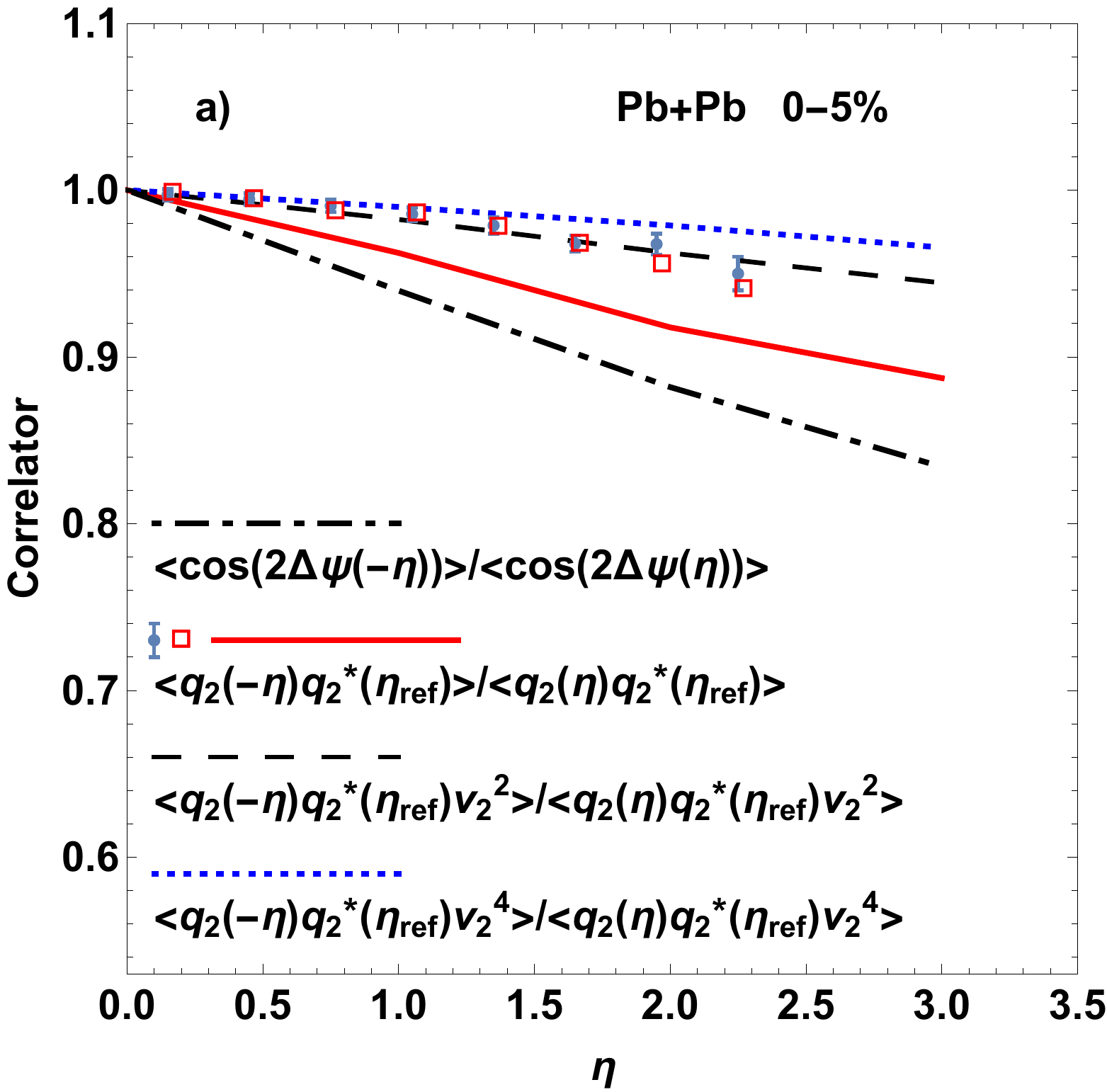}  
\vspace{4mm}

\includegraphics[angle=0,width=0.4 \textwidth]{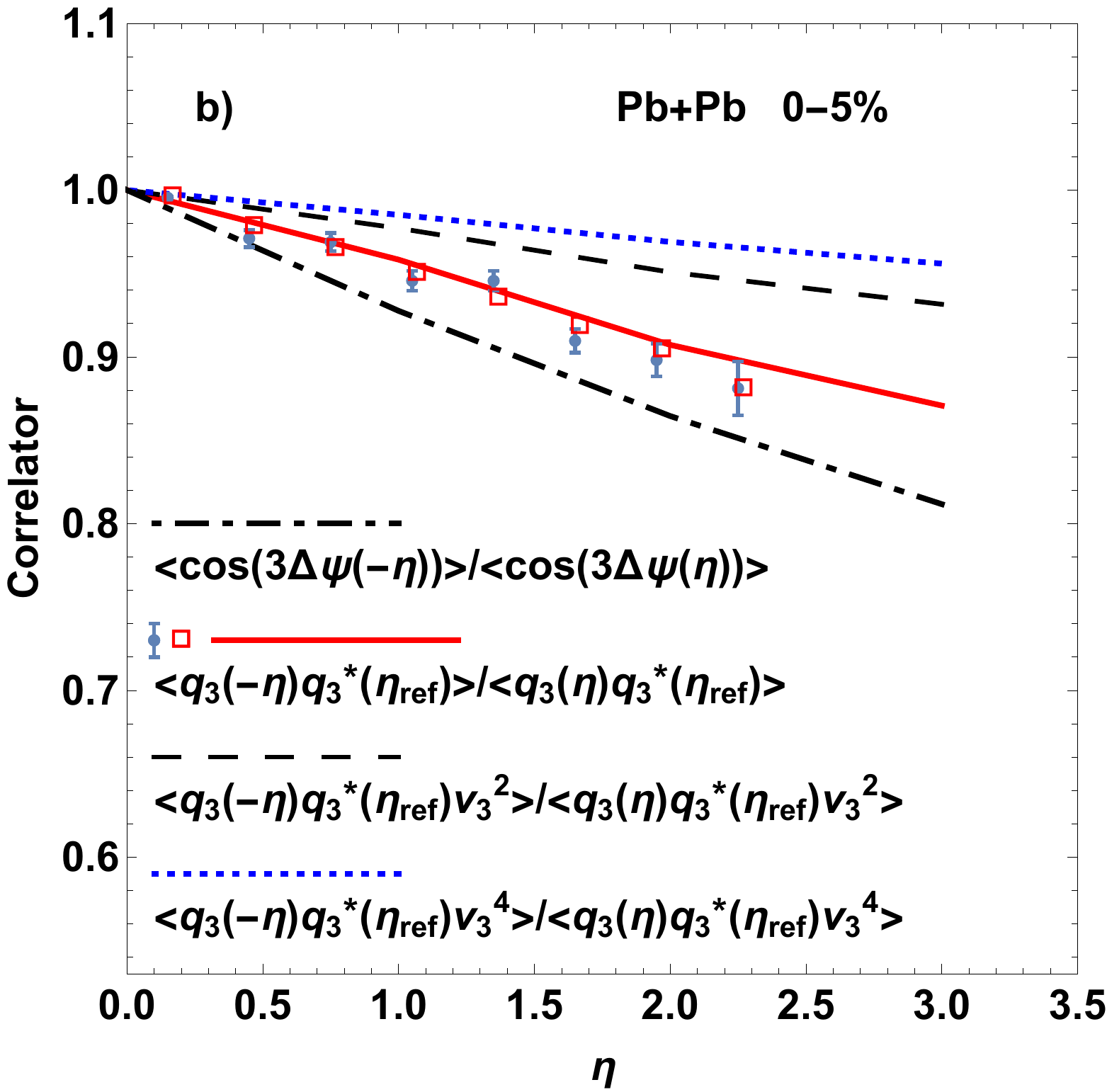}  
\end{center}
\vspace{-5mm}
\caption{Correlator $r_{n|n;1}^{n;2k}$ (Eq. \ref{eq:vk}) for the second- (a) and third-order (b) flow harmonics weighed with flow magnitude 
at mid-rapidity $v_n(\eta=0)^{2k}$, for $k=0$ (solid line), $k=1$ (dashed line), and $k=2$ (dotted line). 
The dashed-dotted line represent  the correlator for 
the flow angle (Eq. \ref{eq:dpsi}). Experimental data for $r_n(\eta)$ are 
represented by diamonds (CMS)  \cite{Khachatryan:2015oea} and empty squares (ATLAS) \cite{Aaboud:2017tql}.
Pb-Pb collisions  for centrality 0-5\%. 
\label{fig:vk05}} 
\end{figure} 

\begin{figure}
\begin{center}
\includegraphics[angle=0,width=0.4 \textwidth]{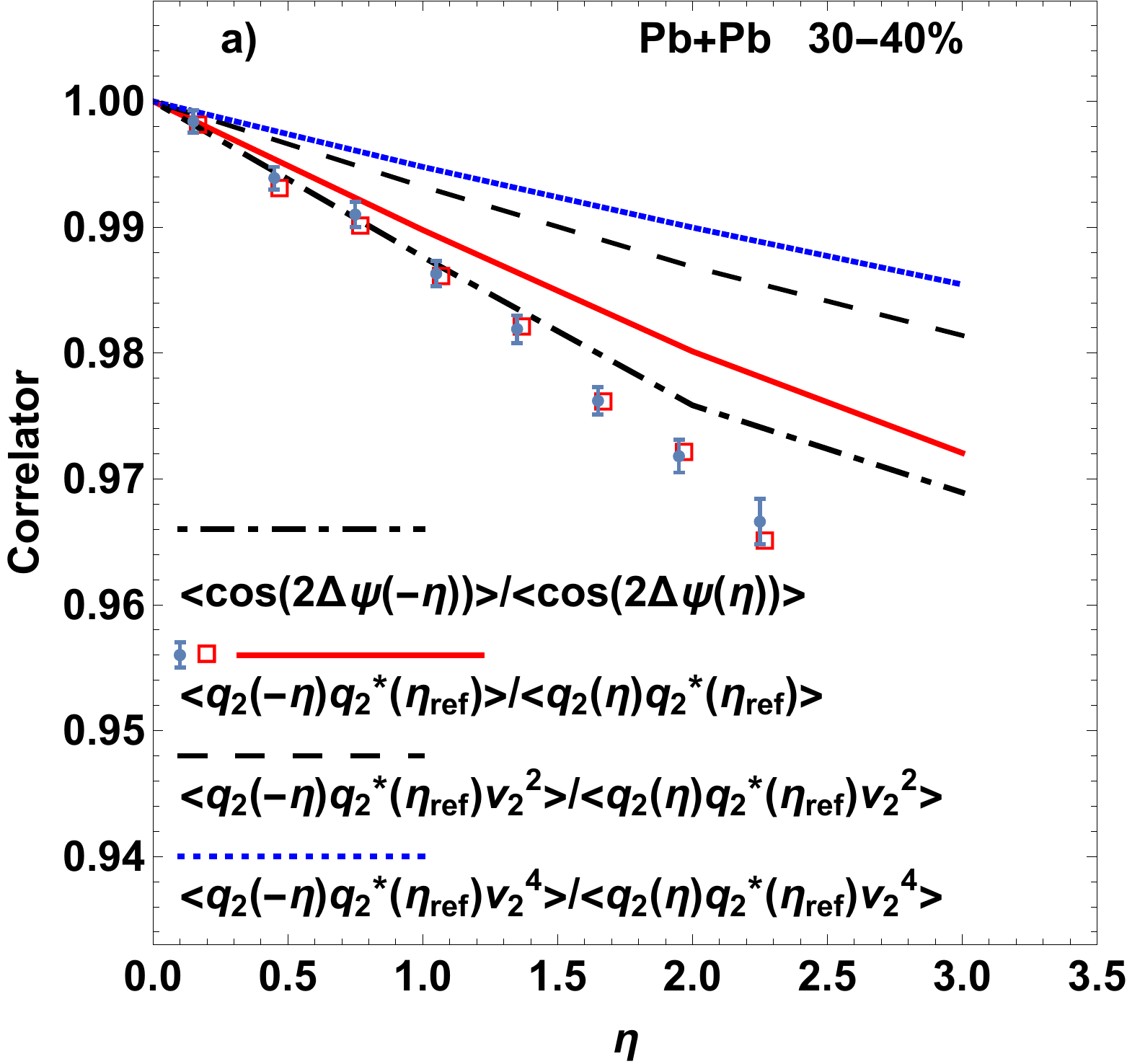}  
\vspace{4mm}

\includegraphics[angle=0,width=0.4 \textwidth]{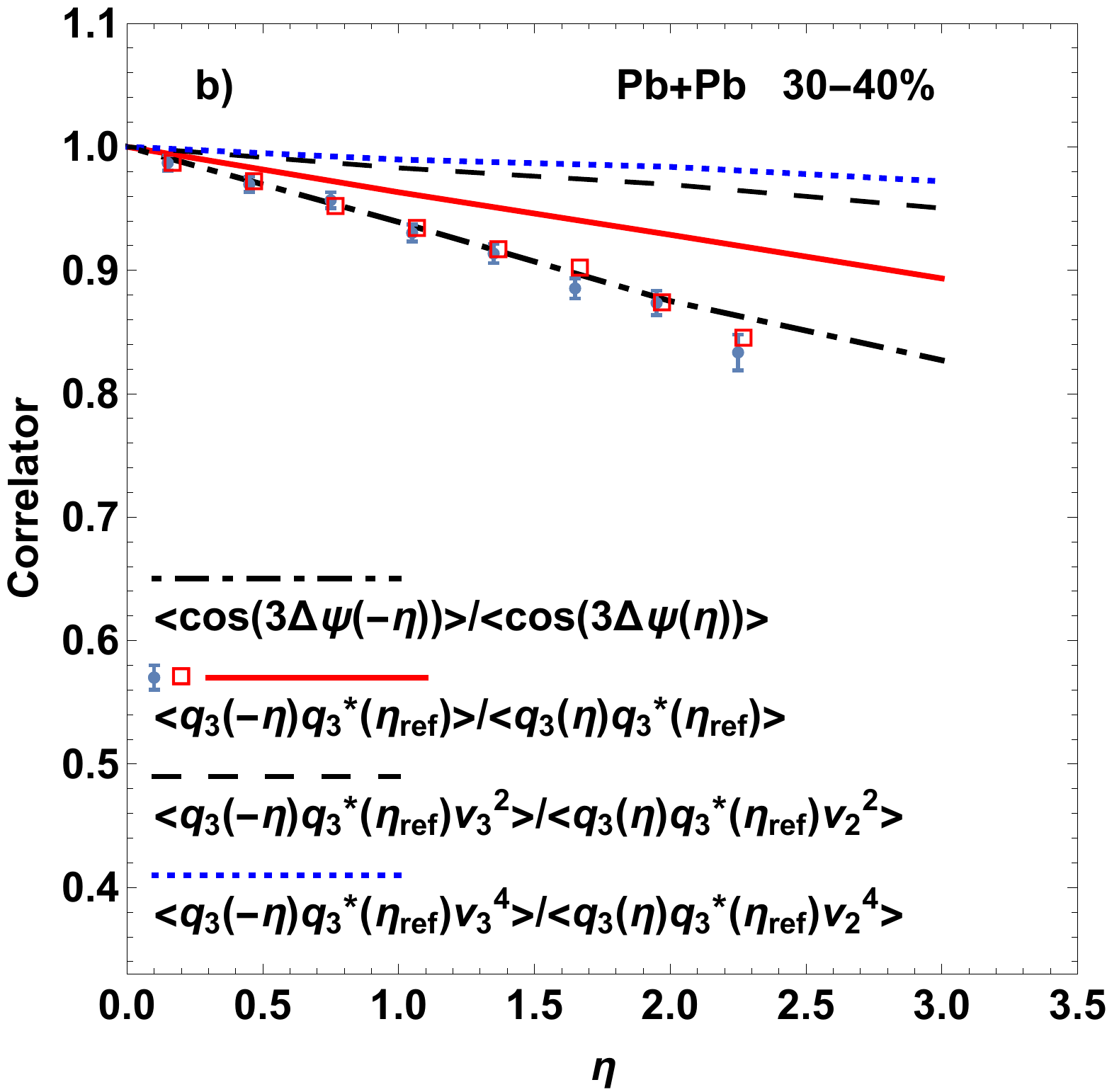}  
\end{center}
\vspace{-5mm}
\caption{Same as Fig. \ref{fig:vk05} but for centrality 30-40\%.
\label{fig:vk3040} }
\end{figure}

Figures \ref{fig:scat05} and \ref{fig:scat3040} present the scattered plots 
of the average  $\cos(n \Delta \Psi )$ and the scaled flow magnitude
$v_n/\langle v_n \rangle$. We note a very specific pattern, which is intuitive: as the flow magnitude increases, i.e., the object is more 
elongated along its principal axis, the forward-backward fluctuations of the flow angle are reduced. We note that  when
$v_n/\langle v_n \rangle \sim 2$, there is hardly any angle fluctuation left. Conversely, when $v_n$ is close to zero, the flow angle is poorly defined, as it 
fluctuates in a wide range.
This effect is visible 
for all the centralities studied, with the flow angle decorrelation largest
for cases where the magnitude of the flow fluctuations is relatively smaller. 
The behavior of the scattered plots (\ref{fig:scat05}-\ref{fig:scat3040}) is qualitatively reflected in 
the features seen in Figs.~(\ref{fig:vv05}-\ref{fig:vv3040}).

The simple understanding of the general behavior is  somewhat toned down by not good agreement of the model results 
with the CMS and ATLAS data, indicated with symbols in Figs.~(\ref{fig:vv05}-\ref{fig:vv3040}), which should agree with 
the solid lines (the flow vector decorrelation). We note too much decorrelation in $r_2$
for the most central events, and too little for $r_2$ and $r_3$ for the mid-peripheral events.  Thus an improvement of the model should 
work in opposite directions for the central and peripheral events. As mentioned in the Introduction, to our knowledge, 
none of the existing models of the flow decorrelations is capable of describing the experimental data in a uniform way.

\section{New measures of  the  correlation between flow magnitude and the torque angle \label{sec:measures}}

\begin{figure}
\begin{center}
\includegraphics[angle=0,width=0.4 \textwidth]{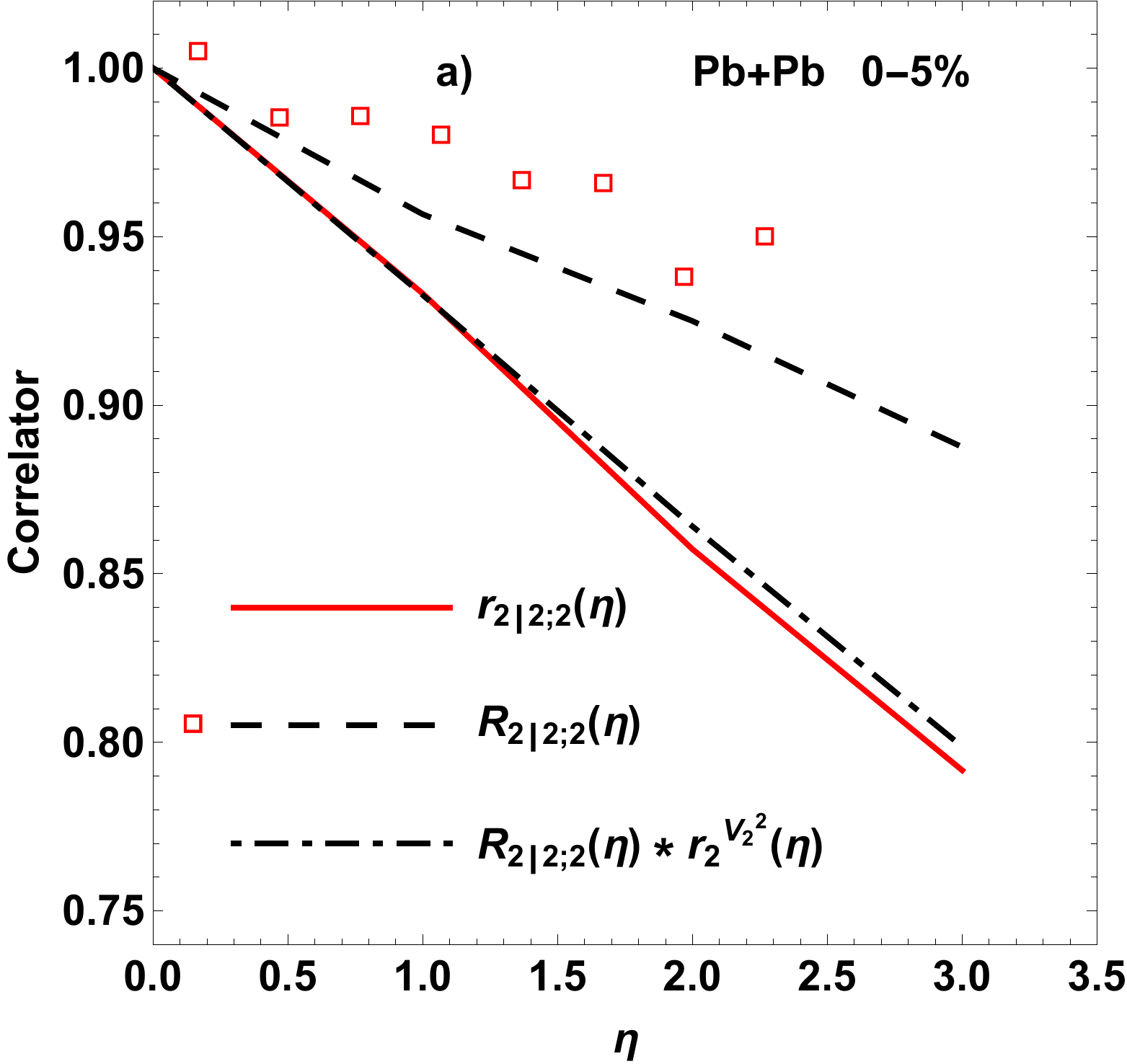}  
\vspace{4mm}

\includegraphics[angle=0,width=0.4 \textwidth]{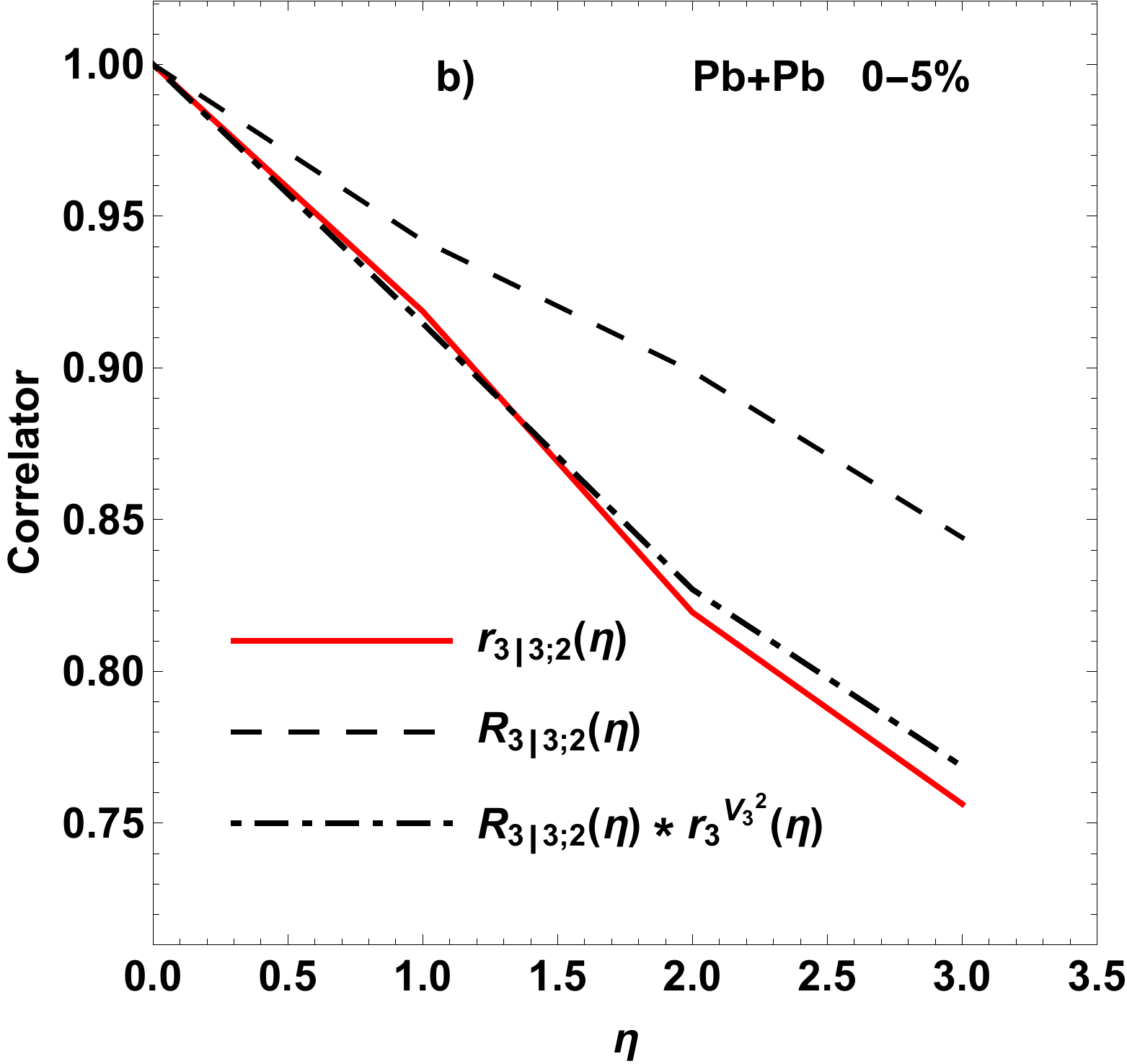}  
\end{center}
\vspace{-5mm}
\caption{The correlators $R_{n|n;2}(\eta)$ (dashed line) and $r_{n|n;2}(\eta)$
for the second- (a) and third-order (b) flow in Pb-Pb collision with centrality 0-5\%.
The ATLAS Collaboration 
data for $R_{2|2;2}$ \cite{Aaboud:2016yar} are represented with empty squares in panel (a).
\label{fig:R05}} 
\end{figure} 

\begin{figure}
\begin{center}
\includegraphics[angle=0,width=0.4 \textwidth]{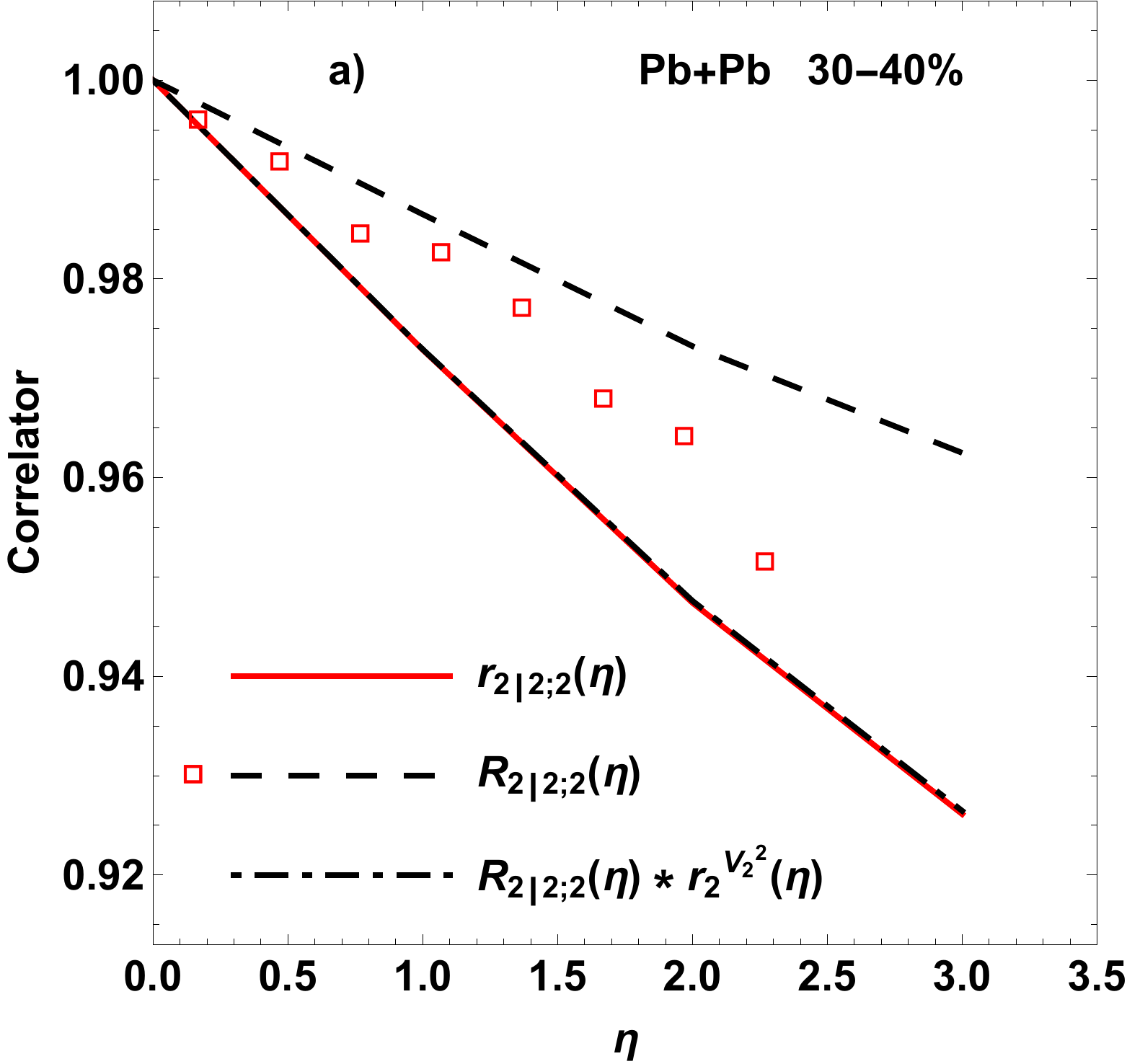}  
\vspace{4mm}

\includegraphics[angle=0,width=0.4 \textwidth]{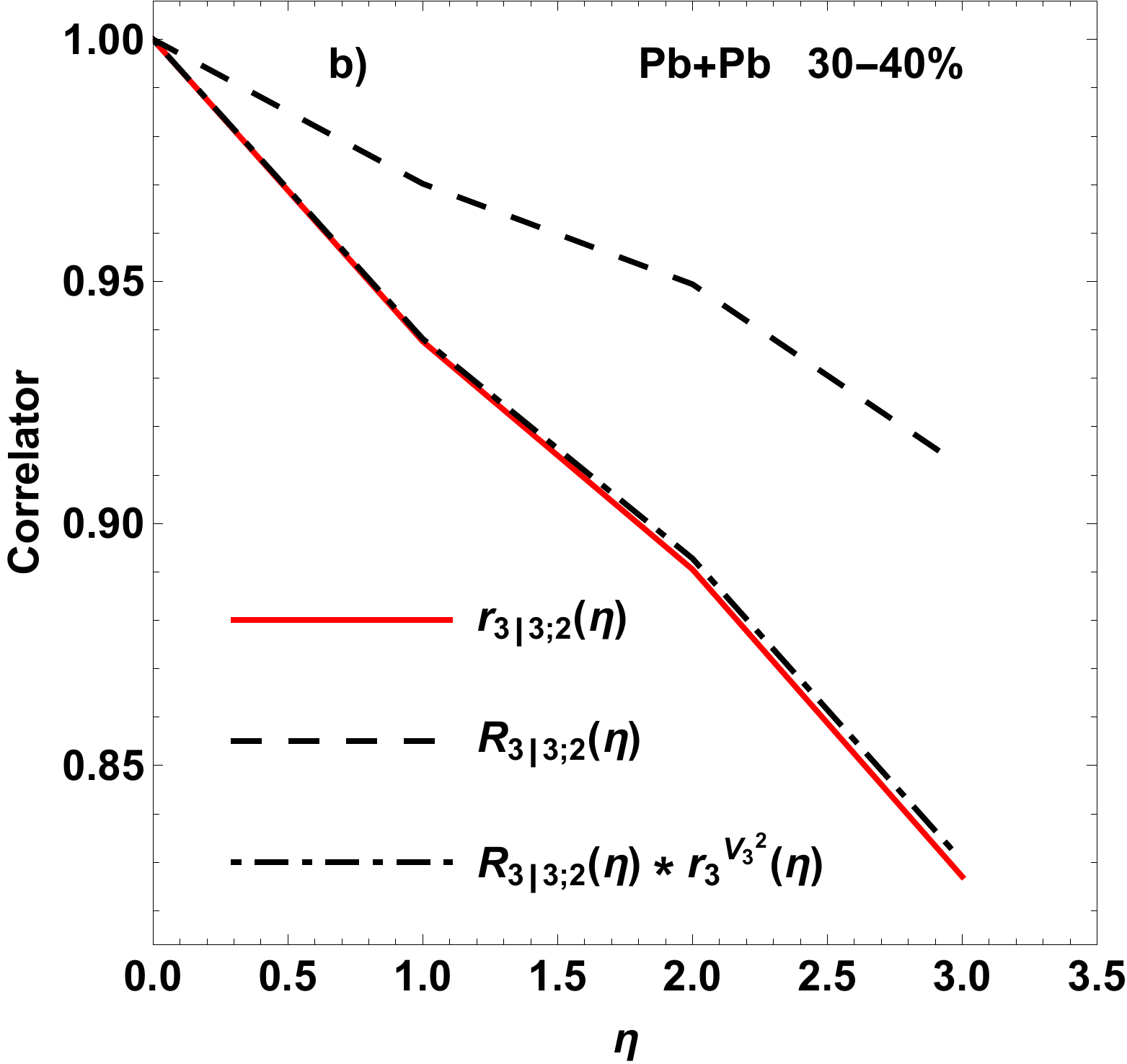}  
\end{center}
\vspace{-5mm}
\caption{Same as Fig. \ref{fig:R05} but for centrality 30-40\%.
\label{fig:R3040} }
\end{figure}

In the observable $r_n(\eta)$, the 
cosine of the torque angle is 
weighted with the second power of the flow magnitude.
The  effect of weighting can be studied more accurately by considering  different
powers of the flow magnitude in the averages such as
\begin{equation}
r_{n|n;1}^{n;2k}(\eta)=\frac{\langle v_n^{2k}(0) q_n(-\eta) q_n^\star(\eta_{ref}) \rangle}{ v_n^{2k}(0) q_n(\eta) q_n^\star(\eta_{ref})}
\label{eq:vk}
\end{equation}
for $k=0,1,2,\dots$ (the factorization breaking coefficient $r_n$ ($=r_{n|n;1})$ of Eq.~(\ref{eq:dvn}) corresponds to $k=0$).

The results of the study of correlations (\ref{eq:vk}) within our hydrodynamic model are shown in Figs.~\ref{fig:vk05} 
and \ref{fig:vk3040}. For correlators involving higher moments of the overall
flow magnitude $v_n^{2k}$, the decorrelation is smaller, 
$ r_{n|n;1}^{n;4}(\eta)>r_{n|n;1}^{n;2}(\eta)>r_{n|n;1}^{n;0}(\eta)$.
The origin of this effect can again be straightforwardly identified with the behavior seen in the scattered plots in Figs.~\ref{fig:scat05} 
and \ref{fig:scat3040}. Higher moments of $v_n$ lead to the preference of higher average
value of $v_n$ in the sampling. Indeed, for $v_3$ and for $v_2$ in central collisions 
one finds from the independent source model the relations 
\begin{eqnarray}
\frac{\langle v_n^2 \rangle^{1/2}}{\langle v_n \rangle} & = &\frac{2}{\sqrt{\pi}} \simeq 1.13, \nonumber \\
\frac{\langle v_n^4 \rangle^{1/4}}{\langle v_n \rangle} & = &\frac{16^{1/4}}{\sqrt{\pi}} \simeq 1.34, \nonumber \\
\frac{\langle v_n^6 \rangle^{1/6}}{\langle v_n \rangle} & = &\frac{96^{1/4}}{\sqrt{\pi}} \simeq 1.52.
\end{eqnarray}
The conclusion at this point is that the hydrodynamic model with the wounded-quark initial condition predicts a strong anti-correlation 
between the flow magnitude and the torque angle in each event.

Similar weighting as in Eq.~(\ref{eq:vk}) could be introduced in other correlators proposed by the ATLAS Collaboration~\cite{Aaboud:2016yar}, e.g.,
\begin{equation}
R_{n|n;2}^{n;2k}(\eta)=\frac{\langle v_n^{2k}(0) q_n(-\eta_{\rm ref})q_n(-\eta) q_n^\star(\eta)q_n^\star(\eta_{\rm ref})\rangle}
{\langle v_n^{2k}(0) q_n(-\eta_{\rm ref})q_n^\star(-\eta) q_n(\eta)q_n^\star(\eta_{\rm ref})\rangle} ,
\end{equation}
or for the nonlinear coupling of the flow harmonics, e.g.,
\begin{equation}
r_{4|4;1}^{2;2k}(\eta)=\frac{\langle v_2^{2k}(0)  q_4(-\eta)q_4^\star(\eta_{\rm ref})\rangle }{\langle v_2^{2k}(0) q_4^\star(\eta) q_4^\star(\eta_{\rm ref}) \rangle} , 
\end{equation}
with $k=0,1,2,\dots$.

\begin{figure}
\begin{center}
\includegraphics[angle=0,width=0.4 \textwidth]{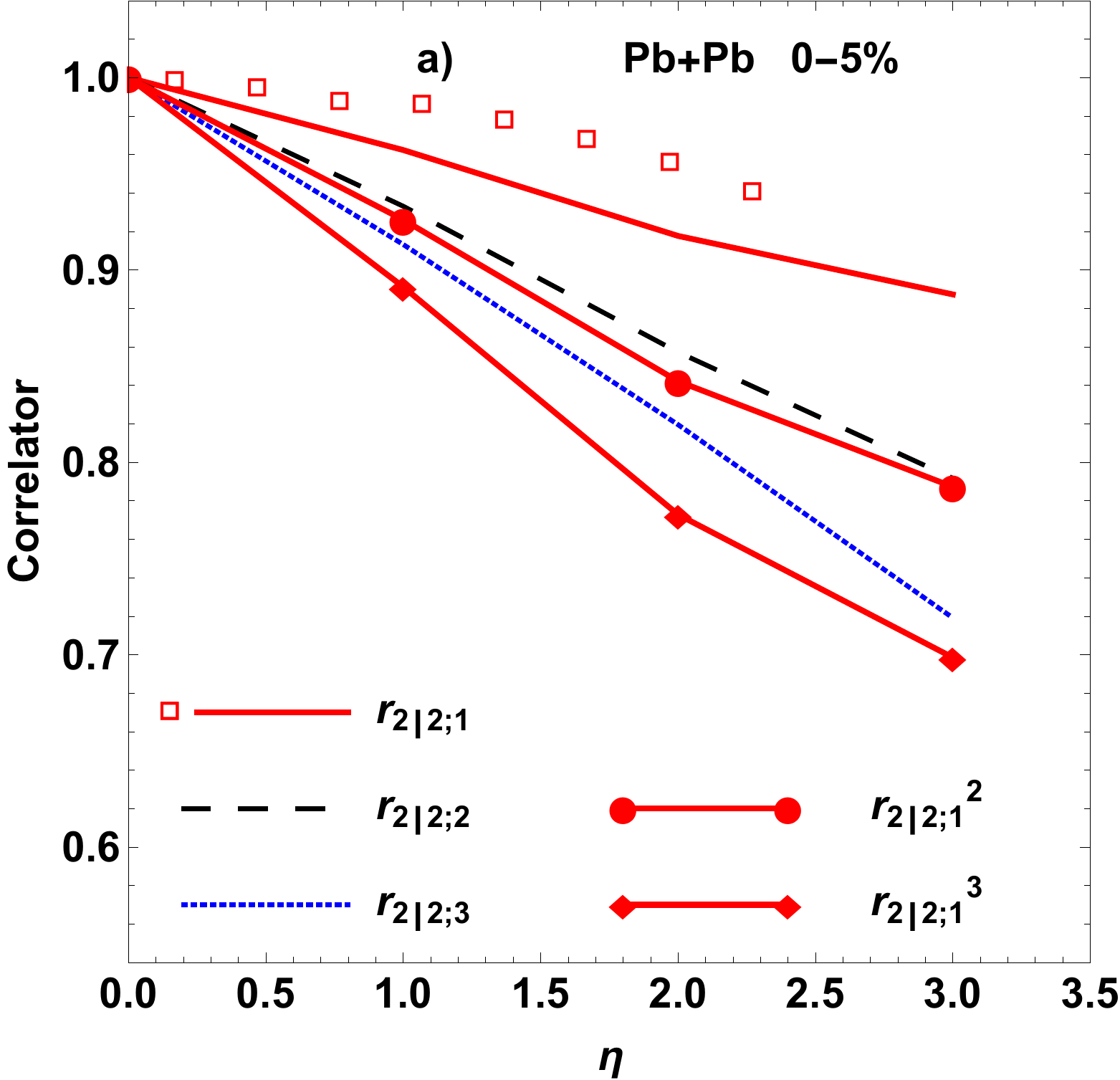}  
\vspace{4mm}

\includegraphics[angle=0,width=0.4 \textwidth]{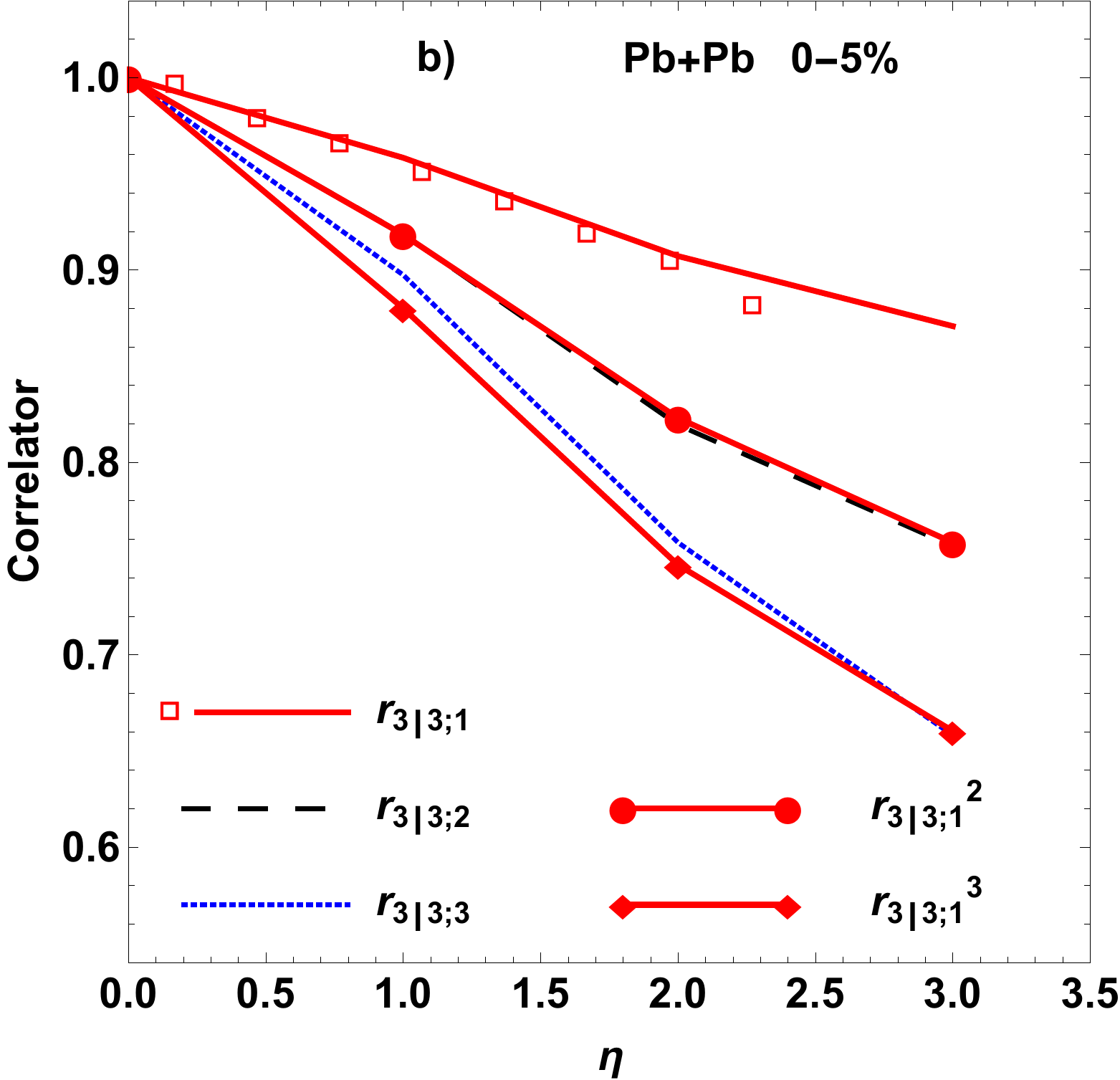}  
\end{center}
\vspace{-5mm}
\caption{Correlators $r_{n|n;k}$ for $k=1,2,3$ in Pb-Pb collisions 
with centrality 0-5\% for the second (panel a) and third (panel b) order harmonic flow.
 Empty squares represent  the ATLAS Collaboration  data \cite{Aaboud:2016yar} 
for $r_{n|n;1}$. 
\label{fig:compare05}} 
\end{figure} 

\begin{figure}
\begin{center}
\includegraphics[angle=0,width=0.4 \textwidth]{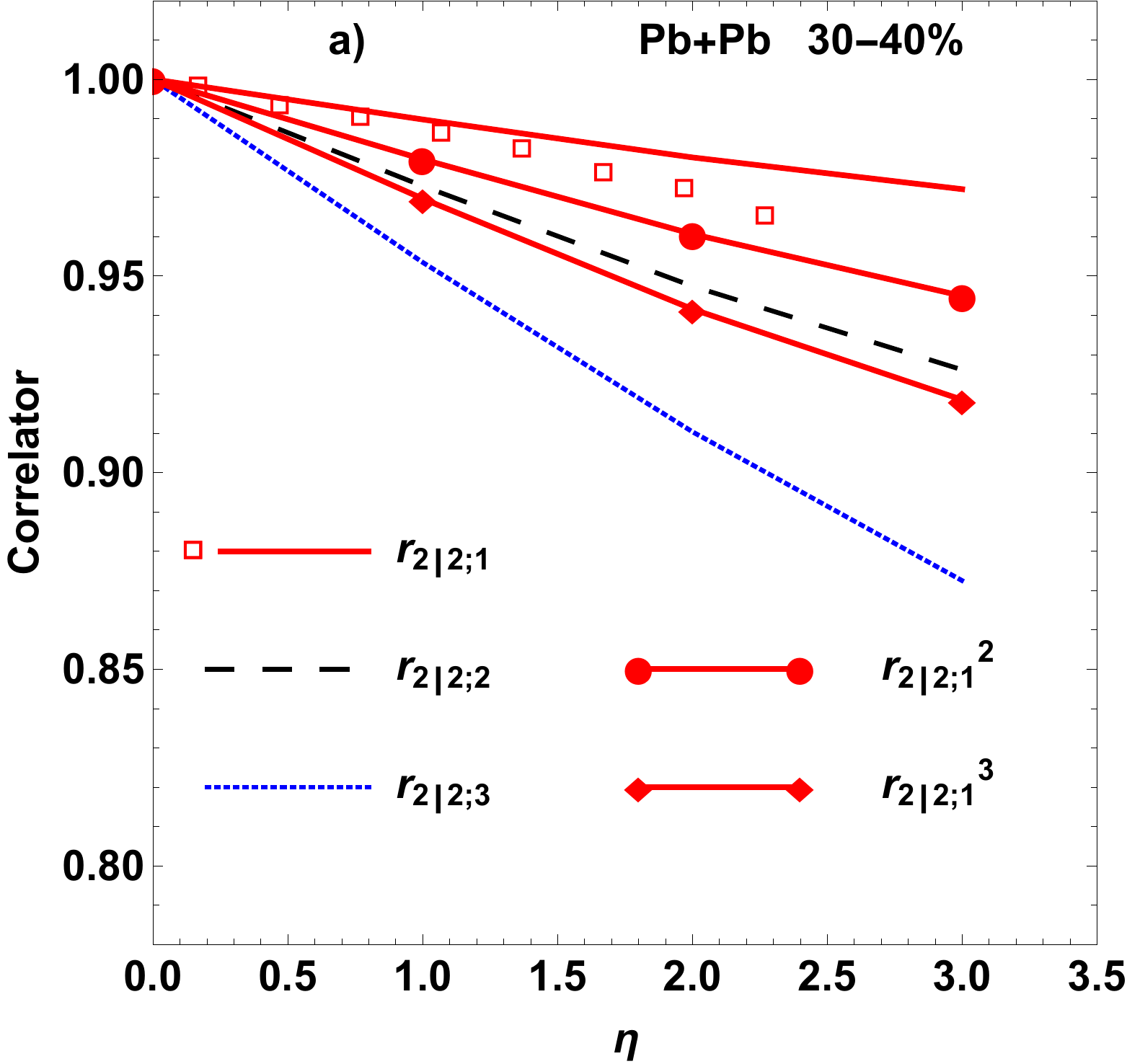}  
\vspace{4mm}

\includegraphics[angle=0,width=0.4 \textwidth]{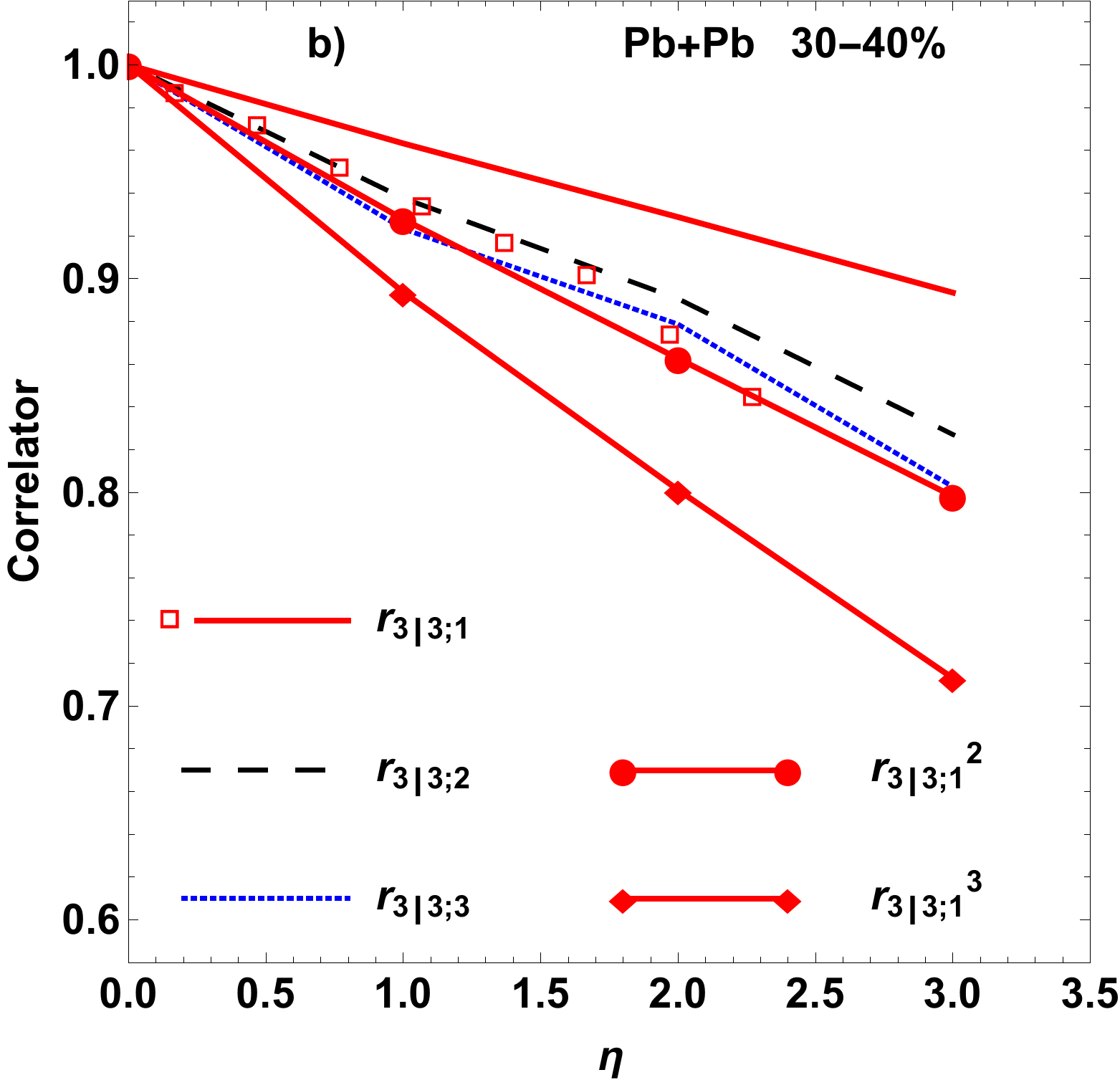}  
\end{center}
\vspace{-5mm}
\caption{Same as Fig. \ref{fig:compare05} but for centrality 30-40\%.
\label{fig:compare3040} }
\end{figure}

\section{Four-bin measures}

The decorrelation of the flow angle at different pseudorapidities can be studied using the 4-bin correlator~\cite{Aaboud:2017tql,Jia:2017kdq} 
\begin{equation}
R_{n|n;2}(\eta)=\frac{ \langle q_n(-\eta_{\rm ref})q_n(-\eta) q_n^\star(\eta)q_n^\star(\eta_{\rm ref})\rangle} 
{ \langle q_n(-\eta_{\rm ref})q_n^\star(-\eta) q_n(\eta)q_n^\star(\eta_{\rm ref})\rangle}   \ .
\end{equation}
It can be compared to the correlators involving a power of the flow vector
in two rapidity bins
\begin{equation}
r_{n|n;k}(\eta)=\frac{\langle q_n^k(-\eta) q_n^{\star\ k}(\eta_{\rm ref})\rangle }{ \langle q_n^k(\eta) q_n^{\star\ k}(\eta_{\rm ref}) \rangle}   \ .
\end{equation}
The correlator $r_{n|n;k}$ measures both the decorrelation  of the 
flow angle and  magnitude $v_n^k$.

Our model results for the correlators $R_{n|n;2}$ and $r_{n|n;2}$ are shown in 
Figs.~\ref{fig:R05} and \ref{fig:R3040}. The agreement with the experimental data 
is qualitatively correct. The decorrelation seen in the measure $r_{n|n|;2}$ is stronger than
in $R_{n|n;2}$. Quantitatively, the model overestimates the flow decorrelation
in central  and underestimates in  semi-peripheral collisions, similarly to the behavior in Figs.~\ref{fig:vv05} and \ref{fig:vv3040}.

There is an interesting feature which we confirm in our simulations:
a factorization of the decorrelation measure $r_{n|n;2}$ into the flow angle decorrelation $R_{n|n;2}$ and 
the flow magnitude decorrelation of $v_n^2$ holds  
in the hydrodynamic model.
We define the correlation measure 
\begin{equation}
r_n^{v^k}(\eta)=\frac{\langle v_n^k(-\eta)v_n^k(\eta_{\rm ref})\rangle}{\langle  v_n^k(\eta)v_n^k(\eta_{\rm ref}) \rangle}  
\label{eq:dv2kmag}
\end{equation}
and find a very good factorization of the flow and magnitude decorrelation, namely 
\begin{equation}
r_{n|n;2}(\eta)\simeq R_{n|n;2}(\eta) *  r_n^{v^2}(\eta),
\end{equation}
holding both in central and semi-peripheral collisions, cf. Figs.~\ref{fig:R05}
and \ref{fig:R3040}.
We have also checked that the decorrelation of the flow magnitude squared approximately factorizes in the hydrodynamic model
\begin{equation}
r_n^{v^2}(\eta) \simeq \left[ r_n^{v}(\eta) \right]^2 \ .
\end{equation}

The following possible factorization of correlators~\cite{Aaboud:2017tql}
\begin{equation}
r_{n|n;k}(\eta) \simeq \left[ r_{n|n;1}(\eta) \right]^k
\end{equation}
can also be tested in our hydrodynamic model.
We find (cf. Figs.~\ref{fig:compare05} and \ref{fig:compare3040}) an
approximate factorization for $v_2$ and $v_3$  in the central collisions,
and partly for $v_3$ in the 30-40\% centrality collisions.
The factorization is broken for the second-order flow in semi-peripheral collisions. These results are qualitatively in agreement with 
the ATLAS Collaboration results at $\sqrt{s}=5020$~TeV~\cite{Aaboud:2017tql}.

\section{Conclusions}

We have analyzed in detail flow decorrelation measures used up to now in experimental studies and in model simulations, 
as well as proposed generalizations of these measures containing different weighting by the flow magnitude. 
The basic purpose of such studies is to check the general ``geometric'' feature present in the models, namely, 
that the events with a higher flow magnitude decorrelate much less in the angle when 
the pseudorapidity location is changed, compared to the events with a lower flow magnitude. 

The generalization of the flow decorrelation measures proposed in this paper, where $v_n^k$ weights are 
additionally incorporated, can be directly tested in experimental analyses.
We have also confirmed a  specific factorization of the angle and magnitude decorrelation between two related correlators, 
as suggested in Ref.~\cite{Aaboud:2017tql}.

One should admit that on the physics side there is a high demand for a deeper understanding of the mechanism generating the early stage fluctuations 
manifest in the forward-backward flow decorrelations.  At this stage of theoretical development, 
the rich experimental data cannot be properly reproduced in a uniform
way, i.e., for all centralities,  orders of the harmonic flow ($n=2,3$), and for various colliding systems (Pb+Pb, p+Pb).

Nonetheless, whereas our model, with the obtained level of agreement to the experimental data, can be viewed as a specific illustrative example only, 
we note that the qualitative agreement with the data has been reached, with the strength of the decorrelation in the right ball-park, 
and with the proper ordering of various correlation measures. In particular, the 
ordering of correlators involving different moments of $v_n$ is the same as in the data.

\section*{Acknowledgments}

Research supported by the Polish Ministry of Science and Higher Education (MNiSW), by the National
Science Centre grant 2015/17/B/ST2/00101 (PB) and grant 2015/19/B/ST2/00937 
(WB), as well as by PL-Grid Infrastructure. Some numerical simulations were carried out in laboratories created under the project
``Development of research base of specialized laboratories of public universities in Swietokrzyskie region'', POIG 02.2.00-26-023/08.
 
\bigskip

\bibliography{../hydr}

\begin{thebibliography}{10}%
\makeatletter
\providecommand \@ifxundefined [1]{%
 \ifx #1\undefined \expandafter \@firstoftwo
 \else \expandafter \@secondoftwo
\fi
}%
\providecommand \@ifnum [1]{%
 \ifnum #1\expandafter \@firstoftwo
 \else \expandafter \@secondoftwo
\fi
}%
\providecommand \enquote [1]{``#1''}%
\providecommand \bibnamefont  [1]{#1}%
\providecommand \bibfnamefont [1]{#1}%
\providecommand \citenamefont [1]{#1}%
\providecommand\href[0]{\@sanitize\@href}%
\providecommand\@href[1]{\endgroup\@@startlink{#1}\endgroup\@@href}%
\providecommand\@@href[1]{#1\@@endlink}%
\providecommand \@sanitize [0]{\begingroup\catcode`\&12\catcode`\#12\relax}%
\@ifxundefined \pdfoutput {\@firstoftwo}{%
 \@ifnum{\z@=\pdfoutput}{\@firstoftwo}{\@secondoftwo}%
}{%
 \providecommand\@@startlink[1]{\leavevmode\special{html:<a href="#1">}}%
 \providecommand\@@endlink[0]{\special{html:</a>}}%
}{%
 \providecommand\@@startlink[1]{%
  \leavevmode
  \pdfstartlink
   attr{/Border[0 0 1 ]/H/I/C[0 1 1]}%
   user{/Subtype/Link/A<</Type/Action/S/URI/URI(#1)>>}%
  \relax
 }%
 \providecommand\@@endlink[0]{\pdfendlink}%
}%
\providecommand \url  [0]{\begingroup\@sanitize \@url }%
\providecommand \@url [1]{\endgroup\@href {#1}{\urlprefix}}%
\providecommand \urlprefix [0]{URL }%
\providecommand \Eprint[0]{\href }%
\@ifxundefined \urlstyle {%
  \providecommand \doi [1]{doi:\discretionary{}{}{}#1}%
}{%
  \providecommand \doi [0]{doi:\discretionary{}{}{}\begingroup
  \urlstyle{rm}\Url }%
}%
\providecommand \doibase [0]{http://dx.doi.org/}%
\providecommand \Doi[1]{\href{\doibase#1}}%
\providecommand \bibAnnote [3]{%
  \BibitemShut{#1}%
  \begin{quotation}\noindent
    \textsc{Key:}\ #2\\\textsc{Annotation:}\ #3%
  \end{quotation}%
}%
\providecommand \bibAnnoteFile [2]{%
  \IfFileExists{#2}{\bibAnnote {#1} {#2} {\input{#2}}}{}%
}%
\providecommand \typeout [0]{\immediate \write \m@ne }%
\providecommand \selectlanguage [0]{\@gobble}%
\providecommand \bibinfo [0]{\@secondoftwo}%
\providecommand \bibfield [0]{\@secondoftwo}%
\providecommand \translation [1]{[#1]}%
\providecommand \BibitemOpen[0]{}%
\providecommand \bibitemStop [0]{}%
\providecommand \bibitemNoStop [0]{.\EOS\space}%
\providecommand \EOS [0]{\spacefactor3000\relax}%
\providecommand \BibitemShut [1]{\csname bibitem#1\endcsname}%
\bibitem{Khachatryan:2015oea}%
  \BibitemOpen
  \bibfield{author}{%
  \bibinfo {author} {\bibfnamefont{V.}~\bibnamefont{Khachatryan}} \emph{et~al.}
  (\bibinfo {collaboration} {CMS Collaboration}),\ }%
  \bibfield{journal}{%
  \Doi{10.1103/PhysRevC.92.034911}{\bibinfo {journal} {Phys. Rev.}}\ }%
  \textbf{\bibinfo {volume} {C92}},\ \bibinfo {pages} {034911} (\bibinfo {year}
  {2015})%
  \bibAnnoteFile{NoStop}{Khachatryan:2015oea}%
\bibitem{Huo:2017hjv}%
  \BibitemOpen
  \bibfield{author}{%
  \bibinfo {author} {\bibfnamefont{P.}~\bibnamefont{Huo}} (\bibinfo
  {collaboration} {ATLAS Collaboration}),\ }%
  \bibfield{journal}{%
  \Doi{10.1016/j.nuclphysa.2017.05.102}{\bibinfo {journal} {Nucl. Phys.}}\ }%
  \textbf{\bibinfo {volume} {A967}},\ \bibinfo {pages} {908} (\bibinfo {year}
  {2017})%
  \bibAnnoteFile{NoStop}{Huo:2017hjv}%
\bibitem{Aaboud:2017tql}%
  \BibitemOpen
  \bibfield{author}{%
  \bibinfo {author} {\bibfnamefont{M.}~\bibnamefont{Aaboud}} \emph{et~al.}
  (\bibinfo {collaboration} {ATLAS Collaboration})}%
   (\bibinfo {year} {2017}),\
  \Eprint{http://arxiv.org/abs/1709.02301}{arXiv:1709.02301 [nucl-ex]}%
  \bibAnnoteFile{NoStop}{Aaboud:2017tql}%
\bibitem{NieTalk}%
  \BibitemOpen
  \bibfield{author}{%
  \bibinfo {author} {\bibfnamefont{M.}~\bibnamefont{Nie}} (\bibinfo
  {collaboration} {STAR Collaboration}),\ }%
  \bibfield{journal}{%
  \bibinfo {journal} {talk presented at 4th International Conference on the
  Initial Stages in High-Energy Nuclear Collisions. Cracow, 18-22 September~}}%
   (\bibinfo {year} {2017})%
  \bibAnnoteFile{NoStop}{NieTalk}%
\bibitem{Bozek:2010vz}%
  \BibitemOpen
  \bibfield{author}{%
  \bibinfo {author} {\bibfnamefont{P.}~\bibnamefont{Bo\.zek}}, \bibinfo
  {author} {\bibfnamefont{W.}~\bibnamefont{Broniowski}},\ and\ \bibinfo
  {author} {\bibfnamefont{J.}~\bibnamefont{Moreira}},\ }%
  \bibfield{journal}{%
  \Doi{10.1103/PhysRevC.83.034911}{\bibinfo {journal} {Phys. Rev.}}\ }%
  \textbf{\bibinfo {volume} {C83}},\ \bibinfo {pages} {034911} (\bibinfo {year}
  {2011})%
  \bibAnnoteFile{NoStop}{Bozek:2010vz}%
\bibitem{Huo:2013qma}%
  \BibitemOpen
  \bibfield{author}{%
  \bibinfo {author} {\bibfnamefont{P.}~\bibnamefont{Huo}}, \bibinfo {author}
  {\bibfnamefont{J.}~\bibnamefont{Jia}},\ and\ \bibinfo {author}
  {\bibfnamefont{S.}~\bibnamefont{Mohapatra}},\ }%
  \bibfield{journal}{%
  \Doi{10.1103/PhysRevC.90.024910}{\bibinfo {journal} {Phys. Rev.}}\ }%
  \textbf{\bibinfo {volume} {C90}},\ \bibinfo {pages} {024910} (\bibinfo {year}
  {2014})%
  \bibAnnoteFile{NoStop}{Huo:2013qma}%
\bibitem{Jia:2014vja}%
  \BibitemOpen
  \bibfield{author}{%
  \bibinfo {author} {\bibfnamefont{J.}~\bibnamefont{Jia}}\ and\ \bibinfo
  {author} {\bibfnamefont{P.}~\bibnamefont{Huo}},\ }%
  \bibfield{journal}{%
  \Doi{10.1103/PhysRevC.90.034905}{\bibinfo {journal} {Phys.Rev.}}\ }%
  \textbf{\bibinfo {volume} {C90}},\ \bibinfo {pages} {034905} (\bibinfo {year}
  {2014})%
  \bibAnnoteFile{NoStop}{Jia:2014vja}%
\bibitem{Jia:2014ysa}%
  \BibitemOpen
  \bibfield{author}{%
  \bibinfo {author} {\bibfnamefont{J.}~\bibnamefont{Jia}}\ and\ \bibinfo
  {author} {\bibfnamefont{P.}~\bibnamefont{Huo}},\ }%
  \bibfield{journal}{%
  \Doi{10.1103/PhysRevC.90.034915}{\bibinfo {journal} {Phys.Rev.}}\ }%
  \textbf{\bibinfo {volume} {C90}},\ \bibinfo {pages} {034915} (\bibinfo {year}
  {2014})%
  \bibAnnoteFile{NoStop}{Jia:2014ysa}%
\bibitem{Bialas:1976ed}%
  \BibitemOpen
  \bibfield{author}{%
  \bibinfo {author} {\bibfnamefont{A.}~\bibnamefont{Bia\l{}as}}, \bibinfo
  {author} {\bibfnamefont{M.}~\bibnamefont{B\l{}eszy\'nski}},\ and\ \bibinfo
  {author} {\bibfnamefont{W.}~\bibnamefont{Czy\.z}},\ }%
  \bibfield{journal}{%
  \bibinfo {journal} {Nucl. Phys.}\ }%
  \textbf{\bibinfo {volume} {B111}},\ \bibinfo {pages} {461} (\bibinfo {year}
  {1976})%
  \bibAnnoteFile{NoStop}{Bialas:1976ed}%
\bibitem{Bozek:2015tca}%
  \BibitemOpen
  \bibfield{author}{%
  \bibinfo {author} {\bibfnamefont{P.}~\bibnamefont{Bozek}}, \bibinfo {author}
  {\bibfnamefont{W.}~\bibnamefont{Broniowski}},\ and\ \bibinfo {author}
  {\bibfnamefont{A.}~\bibnamefont{Olszewski}},\ }%
  \bibfield{journal}{%
  \Doi{10.1103/PhysRevC.92.054913}{\bibinfo {journal} {Phys. Rev.}}\ }%
  \textbf{\bibinfo {volume} {C92}},\ \bibinfo {pages} {054913} (\bibinfo {year}
  {2015})%
  \bibAnnoteFile{NoStop}{Bozek:2015tca}%
\bibitem{Lin:2004en}%
  \BibitemOpen
  \bibfield{author}{%
  \bibinfo {author} {\bibfnamefont{Z.-W.}\ \bibnamefont{Lin}}, \bibinfo
  {author} {\bibfnamefont{C.~M.}\ \bibnamefont{Ko}}, \bibinfo {author}
  {\bibfnamefont{B.-A.}\ \bibnamefont{Li}}, \bibinfo {author}
  {\bibfnamefont{B.}~\bibnamefont{Zhang}},\ and\ \bibinfo {author}
  {\bibfnamefont{S.}~\bibnamefont{Pal}},\ }%
  \bibfield{journal}{%
  \Doi{10.1103/PhysRevC.72.064901}{\bibinfo {journal} {Phys. Rev.}}\ }%
  \textbf{\bibinfo {volume} {C72}},\ \bibinfo {pages} {064901} (\bibinfo {year}
  {2005})%
  \bibAnnoteFile{NoStop}{Lin:2004en}%
\bibitem{Pang:2015zrq}%
  \BibitemOpen
  \bibfield{author}{%
  \bibinfo {author} {\bibfnamefont{L.-G.}\ \bibnamefont{Pang}}, \bibinfo
  {author} {\bibfnamefont{H.}~\bibnamefont{Petersen}}, \bibinfo {author}
  {\bibfnamefont{G.-Y.}\ \bibnamefont{Qin}}, \bibinfo {author}
  {\bibfnamefont{V.}~\bibnamefont{Roy}},\ and\ \bibinfo {author}
  {\bibfnamefont{X.-N.}\ \bibnamefont{Wang}},\ }%
  \bibfield{journal}{%
  \Doi{10.1140/epja/i2016-16097-x}{\bibinfo {journal} {Eur. Phys. J.}}\ }%
  \textbf{\bibinfo {volume} {A52}},\ \bibinfo {pages} {97} (\bibinfo {year}
  {2016})%
  \bibAnnoteFile{NoStop}{Pang:2015zrq}%
\bibitem{Gardim:2012im}%
  \BibitemOpen
  \bibfield{author}{%
  \bibinfo {author} {\bibfnamefont{F.~G.}\ \bibnamefont{Gardim}}, \bibinfo
  {author} {\bibfnamefont{F.}~\bibnamefont{Grassi}}, \bibinfo {author}
  {\bibfnamefont{M.}~\bibnamefont{Luzum}},\ and\ \bibinfo {author}
  {\bibfnamefont{J.-Y.}\ \bibnamefont{Ollitrault}},\ }%
  \bibfield{journal}{%
  \Doi{10.1103/PhysRevC.87.031901}{\bibinfo {journal} {Phys.Rev.}}\ }%
  \textbf{\bibinfo {volume} {C87}},\ \bibinfo {pages} {031901} (\bibinfo {year}
  {2013})%
  \bibAnnoteFile{NoStop}{Gardim:2012im}%
\bibitem{Heinz:2013bua}%
  \BibitemOpen
  \bibfield{author}{%
  \bibinfo {author} {\bibfnamefont{U.}~\bibnamefont{Heinz}}, \bibinfo {author}
  {\bibfnamefont{Z.}~\bibnamefont{Qiu}},\ and\ \bibinfo {author}
  {\bibfnamefont{C.}~\bibnamefont{Shen}},\ }%
  \bibfield{journal}{%
  \Doi{10.1103/PhysRevC.87.034913}{\bibinfo {journal} {Phys.Rev.}}\ }%
  \textbf{\bibinfo {volume} {C87}},\ \bibinfo {pages} {034913} (\bibinfo {year}
  {2013})%
  \bibAnnoteFile{NoStop}{Heinz:2013bua}%
\bibitem{Kozlov:2014fqa}%
  \BibitemOpen
  \bibfield{author}{%
  \bibinfo {author} {\bibfnamefont{I.}~\bibnamefont{Kozlov}}, \bibinfo {author}
  {\bibfnamefont{M.}~\bibnamefont{Luzum}}, \bibinfo {author}
  {\bibfnamefont{G.}~\bibnamefont{Denicol}}, \bibinfo {author}
  {\bibfnamefont{S.}~\bibnamefont{Jeon}},\ and\ \bibinfo {author}
  {\bibfnamefont{C.}~\bibnamefont{Gale}}}%
   (\bibinfo {year} {2014}),\
  \Eprint{http://arxiv.org/abs/1405.3976}{arXiv:1405.3976 [nucl-th]}%
  \bibAnnoteFile{NoStop}{Kozlov:2014fqa}%
\bibitem{Acharya:2017ino}%
  \BibitemOpen
  \bibfield{author}{%
  \bibinfo {author} {\bibfnamefont{S.}~\bibnamefont{Acharya}} \emph{et~al.}
  (\bibinfo {collaboration} {ALICE Collaboration}),\ }%
  \bibfield{journal}{%
  \Doi{10.1007/JHEP09(2017)032}{\bibinfo {journal} {JHEP}}\ }%
  \textbf{\bibinfo {volume} {09}},\ \bibinfo {pages} {032} (\bibinfo {year}
  {2017})%
  \bibAnnoteFile{NoStop}{Acharya:2017ino}%
\bibitem{Aad:2014lta}%
  \BibitemOpen
  \bibfield{author}{%
  \bibinfo {author} {\bibfnamefont{G.}~\bibnamefont{Aad}} \emph{et~al.}
  (\bibinfo {collaboration} {ATLAS Collaboration}),\ }%
  \bibfield{journal}{%
  \Doi{10.1103/PhysRevC.90.044906}{\bibinfo {journal} {Phys.Rev.}}\ }%
  \textbf{\bibinfo {volume} {C90}},\ \bibinfo {pages} {044906} (\bibinfo {year}
  {2014})%
  \bibAnnoteFile{NoStop}{Aad:2014lta}%
\bibitem{Bozek:2015bna}%
  \BibitemOpen
  \bibfield{author}{%
  \bibinfo {author} {\bibfnamefont{P.}~\bibnamefont{Bo{\.z}ek}}\ and\ \bibinfo
  {author} {\bibfnamefont{W.}~\bibnamefont{Broniowski}},\ }%
  \bibfield{journal}{%
  \Doi{10.1016/j.physletb.2015.11.054}{\bibinfo {journal} {Phys. Lett.}}\ }%
  \textbf{\bibinfo {volume} {B752}},\ \bibinfo {pages} {206} (\bibinfo {year}
  {2016})%
  \bibAnnoteFile{NoStop}{Bozek:2015bna}%
\bibitem{Broniowski:2015oif}%
  \BibitemOpen
  \bibfield{author}{%
  \bibinfo {author} {\bibfnamefont{W.}~\bibnamefont{Broniowski}}\ and\ \bibinfo
  {author} {\bibfnamefont{P.}~\bibnamefont{Bożek}},\ }%
  \bibfield{journal}{%
  \Doi{10.1103/PhysRevC.93.064910}{\bibinfo {journal} {Phys. Rev.}}\ }%
  \textbf{\bibinfo {volume} {C93}},\ \bibinfo {pages} {064910} (\bibinfo {year}
  {2016})%
  \bibAnnoteFile{NoStop}{Broniowski:2015oif}%
\bibitem{Monnai:2015sca}%
  \BibitemOpen
  \bibfield{author}{%
  \bibinfo {author} {\bibfnamefont{A.}~\bibnamefont{Monnai}}\ and\ \bibinfo
  {author} {\bibfnamefont{B.}~\bibnamefont{Schenke}},\ }%
  \bibfield{journal}{%
  \Doi{10.1016/j.physletb.2015.11.063}{\bibinfo {journal} {Phys. Lett.}}\ }%
  \textbf{\bibinfo {volume} {B752}},\ \bibinfo {pages} {317} (\bibinfo {year}
  {2016})%
  \bibAnnoteFile{NoStop}{Monnai:2015sca}%
\bibitem{Sakai:2017rfi}%
  \BibitemOpen
  \bibfield{author}{%
  \bibinfo {author} {\bibfnamefont{A.}~\bibnamefont{Sakai}}, \bibinfo {author}
  {\bibfnamefont{K.}~\bibnamefont{Murase}},\ and\ \bibinfo {author}
  {\bibfnamefont{T.}~\bibnamefont{Hirano}},\ }%
  \bibfield{journal}{%
  \Doi{10.1016/j.nuclphysa.2017.05.010}{\bibinfo {journal} {Nucl. Phys.}}\ }%
  \textbf{\bibinfo {volume} {A967}},\ \bibinfo {pages} {445} (\bibinfo {year}
  {2017})%
  \bibAnnoteFile{NoStop}{Sakai:2017rfi}%
\bibitem{Schenke:2016ksl}%
  \BibitemOpen
  \bibfield{author}{%
  \bibinfo {author} {\bibfnamefont{B.}~\bibnamefont{Schenke}}\ and\ \bibinfo
  {author} {\bibfnamefont{S.}~\bibnamefont{Schlichting}},\ }%
  \bibfield{journal}{%
  \Doi{10.1103/PhysRevC.94.044907}{\bibinfo {journal} {Phys. Rev.}}\ }%
  \textbf{\bibinfo {volume} {C94}},\ \bibinfo {pages} {044907} (\bibinfo {year}
  {2016})%
  \bibAnnoteFile{NoStop}{Schenke:2016ksl}%
\bibitem{Shen:2017bsr}%
  \BibitemOpen
  \bibfield{author}{%
  \bibinfo {author} {\bibfnamefont{C.}~\bibnamefont{Shen}}\ and\ \bibinfo
  {author} {\bibfnamefont{B.}~\bibnamefont{Schenke}}}%
   (\bibinfo {year} {2017}),\
  \Eprint{http://arxiv.org/abs/1710.00881}{arXiv:1710.00881 [nucl-th]}%
  \bibAnnoteFile{NoStop}{Shen:2017bsr}%
\bibitem{Ke:2016jrd}%
  \BibitemOpen
  \bibfield{author}{%
  \bibinfo {author} {\bibfnamefont{W.}~\bibnamefont{Ke}}, \bibinfo {author}
  {\bibfnamefont{J.~S.}\ \bibnamefont{Moreland}}, \bibinfo {author}
  {\bibfnamefont{J.~E.}\ \bibnamefont{Bernhard}},\ and\ \bibinfo {author}
  {\bibfnamefont{S.~A.}\ \bibnamefont{Bass}},\ }%
  \bibfield{journal}{%
  \Doi{10.1103/PhysRevC.96.044912}{\bibinfo {journal} {Phys. Rev.}}\ }%
  \textbf{\bibinfo {volume} {C96}},\ \bibinfo {pages} {044912} (\bibinfo {year}
  {2017})%
  \bibAnnoteFile{NoStop}{Ke:2016jrd}%
\bibitem{Bozek:2009dw}%
  \BibitemOpen
  \bibfield{author}{%
  \bibinfo {author} {\bibfnamefont{P.}~\bibnamefont{Bo\.zek}},\ }%
  \bibfield{journal}{%
  \bibinfo {journal} {Phys. Rev.}\ }%
  \textbf{\bibinfo {volume} {C81}},\ \bibinfo {pages} {034909} (\bibinfo {year}
  {2010})%
  \bibAnnoteFile{NoStop}{Bozek:2009dw}%
\bibitem{Schenke:2010rr}%
  \BibitemOpen
  \bibfield{author}{%
  \bibinfo {author} {\bibfnamefont{B.}~\bibnamefont{Schenke}}, \bibinfo
  {author} {\bibfnamefont{S.}~\bibnamefont{Jeon}},\ and\ \bibinfo {author}
  {\bibfnamefont{C.}~\bibnamefont{Gale}},\ }%
  \bibfield{journal}{%
  \Doi{10.1103/PhysRevLett.106.042301}{\bibinfo {journal} {Phys. Rev. Lett.}}\
  }%
  \textbf{\bibinfo {volume} {106}},\ \bibinfo {pages} {042301} (\bibinfo {year}
  {2011})%
  \bibAnnoteFile{NoStop}{Schenke:2010rr}%
\bibitem{Rybczynski:2013yba}%
  \BibitemOpen
  \bibfield{author}{%
  \bibinfo {author} {\bibfnamefont{M.}~\bibnamefont{Rybczy\'nski}}, \bibinfo
  {author} {\bibfnamefont{G.}~\bibnamefont{Stefanek}}, \bibinfo {author}
  {\bibfnamefont{W.}~\bibnamefont{Broniowski}},\ and\ \bibinfo {author}
  {\bibfnamefont{P.}~\bibnamefont{Bo\.zek}},\ }%
  \bibfield{journal}{%
  \Doi{10.1016/j.cpc.2014.02.016}{\bibinfo {journal} {Comput. Phys. Commun.}}\
  }%
  \textbf{\bibinfo {volume} {185}},\ \bibinfo {pages} {1759} (\bibinfo {year}
  {2014})%
  \bibAnnoteFile{NoStop}{Rybczynski:2013yba}%
\bibitem{Bozek:2016kpf}%
  \BibitemOpen
  \bibfield{author}{%
  \bibinfo {author} {\bibfnamefont{P.}~\bibnamefont{Bo{\.z}ek}}, \bibinfo
  {author} {\bibfnamefont{W.}~\bibnamefont{Broniowski}},\ and\ \bibinfo
  {author} {\bibfnamefont{M.}~\bibnamefont{Rybczy{\'n}ski}},\ }%
  \bibfield{journal}{%
  \Doi{10.1103/PhysRevC.94.014902}{\bibinfo {journal} {Phys. Rev.}}\ }%
  \textbf{\bibinfo {volume} {C94}},\ \bibinfo {pages} {014902} (\bibinfo {year}
  {2016})%
  \bibAnnoteFile{NoStop}{Bozek:2016kpf}%
\bibitem{Chojnacki:2011hb}%
  \BibitemOpen
  \bibfield{author}{%
  \bibinfo {author} {\bibfnamefont{M.}~\bibnamefont{Chojnacki}}, \bibinfo
  {author} {\bibfnamefont{A.}~\bibnamefont{Kisiel}}, \bibinfo {author}
  {\bibfnamefont{W.}~\bibnamefont{Florkowski}},\ and\ \bibinfo {author}
  {\bibfnamefont{W.}~\bibnamefont{Broniowski}},\ }%
  \bibfield{journal}{%
  \Doi{10.1016/j.cpc.2011.11.018}{\bibinfo {journal} {Comput. Phys. Commun.}}\
  }%
  \textbf{\bibinfo {volume} {183}},\ \bibinfo {pages} {746} (\bibinfo {year}
  {2012})%
  \bibAnnoteFile{NoStop}{Chojnacki:2011hb}%
\bibitem{Bialas:2004su}%
  \BibitemOpen
  \bibfield{author}{%
  \bibinfo {author} {\bibfnamefont{A.}~\bibnamefont{Bia\l{}as}}\ and\ \bibinfo
  {author} {\bibfnamefont{W.}~\bibnamefont{Czy\.z}},\ }%
  \bibfield{journal}{%
  \bibinfo {journal} {Acta Phys. Polon.}\ }%
  \textbf{\bibinfo {volume} {B36}},\ \bibinfo {pages} {905} (\bibinfo {year}
  {2005})%
  \bibAnnoteFile{NoStop}{Bialas:2004su}%
\bibitem{Bozek:2015bha}%
  \BibitemOpen
  \bibfield{author}{%
  \bibinfo {author} {\bibfnamefont{P.}~\bibnamefont{Bo\.zek}}, \bibinfo
  {author} {\bibfnamefont{W.}~\bibnamefont{Broniowski}},\ and\ \bibinfo
  {author} {\bibfnamefont{A.}~\bibnamefont{Olszewski}},\ }%
  \bibfield{journal}{%
  \Doi{10.1103/PhysRevC.91.054912}{\bibinfo {journal} {Phys.Rev.}}\ }%
  \textbf{\bibinfo {volume} {C91}},\ \bibinfo {pages} {054912} (\bibinfo {year}
  {2015})%
  \bibAnnoteFile{NoStop}{Bozek:2015bha}%
\bibitem{Aaboud:2016yar}%
  \BibitemOpen
  \bibfield{author}{%
  \bibinfo {author} {\bibfnamefont{M.}~\bibnamefont{Aaboud}} \emph{et~al.}
  (\bibinfo {collaboration} {ATLAS}),\ }%
  \bibfield{journal}{%
  \Doi{10.1103/PhysRevC.96.024908}{\bibinfo {journal} {Phys. Rev.}}\ }%
  \textbf{\bibinfo {volume} {C96}},\ \bibinfo {pages} {024908} (\bibinfo {year}
  {2017})%
  \bibAnnoteFile{NoStop}{Aaboud:2016yar}%
\bibitem{Jia:2017kdq}%
  \BibitemOpen
  \bibfield{author}{%
  \bibinfo {author} {\bibfnamefont{J.}~\bibnamefont{Jia}}, \bibinfo {author}
  {\bibfnamefont{P.}~\bibnamefont{Huo}}, \bibinfo {author}
  {\bibfnamefont{G.}~\bibnamefont{Ma}},\ and\ \bibinfo {author}
  {\bibfnamefont{M.}~\bibnamefont{Nie}},\ }%
  \bibfield{journal}{%
  \Doi{10.1088/1361-6471/aa74c3}{\bibinfo {journal} {J. Phys.}}\ }%
  \textbf{\bibinfo {volume} {G44}},\ \bibinfo {pages} {075106} (\bibinfo {year}
  {2017})%
  \bibAnnoteFile{NoStop}{Jia:2017kdq}%
\end{thebibliography}%

\end{document}